\documentclass[letterpaper,11pt]{scrartcl}

\usepackage{paper}

\title{BCL::MP-Fold: membrane protein structure prediction guided by EPR restraints}
\author[1,*]{Axel W. Fischer}
\author[1,*]{Nathan S. Alexander}
\author[1]{Nils Woetzel}
\author[1]{Mert Karakaş}
\author[1]{Brian E. Weiner}
\author[1]{Jens Meiler}
\affil[1]{Department of Chemistry and Center for Structural Biology, Vanderbilt
  University, Nashville, USA}
\affil[*]{Contributed equally to this article}

\date{}

\begin{document}

\maketitle

\begin{abstract}

  For many membrane proteins, the determination of their topology remains a challenge for methods
  like X-ray crystallography and \gls{nmr} spectroscopy. \Gls{epr} spectroscopy has evolved as an
  alternative technique to study structure and dynamics of membrane proteins. The present study
  demonstrates the feasibility of membrane protein topology determination using limited \gls{epr}
  distance and accessibility measurements. The BCL::MP-Fold algorithm assembles \glspl{sse} in the
  membrane using a \gls{mcm} approach. Sampled models are evaluated using knowledge-based potential
  functions and agreement with the \gls{epr} data and a knowledge-based energy
  function. Twenty-nine membrane proteins of up to 696 residues are used to test the algorithm. The
  \gls{rmsd100} value of the most accurate model is better than \SI{8}{\angstrom} for twenty-seven,
  better than \SI{6}{\angstrom} for twenty-two, and better than \SI{4}{\angstrom} for fifteen out
  of twenty-nine proteins, demonstrating the algorithm's ability to sample the native topology. The
  average enrichment could be improved from \num{1.3} to \num{2.5}, showing the improved
  discrimination power by using \gls{epr} data.

\end{abstract}

\section{Introduction}

Membrane protein structure determination continues to be a challenge. About \SI{22}{\percent} of
all proteins are membrane proteins and an estimated \SI{60}{\percent} of pharmaceutical therapies
target membrane proteins \autocite{Overington2006}. However, only \SI{2.5}{\percent} of the
proteins deposited in the \gls{pdb} are classified as membrane proteins \autocite{Tusnady2004,
  Berman2000}. Protein structures are typically determined to atomic detail using X-ray
crystallography or \gls{nmr} spectroscopy. However, membrane proteins provide challenges for both
techniques \autocite{Bill2011}. It is difficult to obtain quantities of purified membrane proteins
sufficient for both X-ray crystallography and \gls{nmr} spectroscopy. The two-dimensional nature of
the membrane complicates crystallization in a three-dimensional crystal lattice. In order to obtain
crystals, the target protein is often subjected to non-native-like environments and/or
modifications such as stabilizing sequence mutations \autocite{Tate2009, Tate2010}. Additional
problems may evolve from post-translational modification such as phosphorylation
\autocite{Kobilka2007}. Many membrane proteins continue to be too large for structure determination
by \gls{nmr} spectroscopy \autocite{Kang2011}. Even if the target itself is not too large, the
membrane mimic adds significant additional mass to the system \autocite{Kim2009}. Despite wonderful
successes in determining the structure of high-profile targets, it is critical that the structural
features observed with one technique are confirmed with an orthogonal technique
\autocite{Alexander2014}.

\Gls{epr} spectroscopy in conjunction with \gls{sdsl} provides such an orthogonal technique for
probing structural aspects of membrane proteins \autocite{Hubbell1994, Dong2005,
  Czogalla2007}. Advantages of \gls{epr} spectroscopy include that the protein can be studied in a
native-like environment and that only a relatively small sample amount is required. In addition,
\gls{epr} spectroscopy can be used to study large proteins. Although \gls{epr} is a versatile tool
for probing membrane protein structure, it has its own challenges: at least one unpaired electron
(spin label) needs to be introduced into the protein. Typically, this requires mutation of all
cysteine residues to either alanine or serine, introduction of one or two cysteines at the desired
labeling sites, coupling to the thiol-specific nitroxide spin label
\textit{S}-(1-oxyl-2,2,5,5-tetramethyl-2,5-dihydro-1H-pyrrol-3-yl)methyl
  methanesulfonothioate (MTSL), and functional characterization of the protein. As a result, data
sets from \gls{epr} spectroscopy are sparse containing only a fraction of measurements per residue
in the target protein. \Gls{epr} is not a high-throughput technique.

\Gls{epr} provides two categories of structural information important to membrane protein
topology: \begin{inparaenum}[a)] \item \Gls{epr} can provide information about the local
environment of the spin label \autocite{Koteiche1998, Koteiche1999, Altenbach1990}. The
accessibility of the spin label to oxygen probe molecules indicates the degree of burial of the
spin label within the protein in the transmembrane region. Accessibility measurements are
typically performed in a sequence scanning fashion. This provides an accessibility profile over a
large portion of the sequence \autocite{Lietzow2004, Altenbach1996}. The accessibility profile
tracks the periodicity of \glspl{sse} as individual measurements rise and fall according to the
periodic exposure and burial of residues. The exposed face of a \gls{sse} can be determined
\autocite{Salwinski1999}, a task that is difficult within the hydrophobic environment of the
membrane. \item When two spin labels are introduced, \gls{epr} can measure inter-spin label
distances, routinely of up to \SI{60}{\angstrom} through the \gls{deer} experiment
\autocite{Borbat2002, Jeschke2007}. \Gls{epr} distance measurements have been demonstrated on
several large membrane proteins including MsbA \autocite{Zou2009a}, rhodopsin
\autocite{Altenbach2008}, and LeuT \autocite{Claxton2010}. \end{inparaenum} Given the sparseness
of data, \gls{epr} has been frequently used to probe different structural states of proteins
\autocite{Chakrapani2010, Vasquez2008}. Changes in distances and accessibilities track regions of
the protein that move when converting from one state into another. Such investigations rely upon an
already determined experimental structure to define the protein topology and provide a scaffold to
map changes observed via \gls{epr} spectroscopy.

\begin{wraptable}{R}{0.53\textwidth}
  \centering
  \small
  \begin{tabular}{lrrrrrr}
\toprule
Protein & \#aas & \#SSE & \si{\percent}$\mathit{res}_{\mathit{SSE}}$ & source & res. \\
\midrule
1IWG  & \num{68}   &\num{5}     & \SI{90}{\percent}        & X-ray & \SI{3.5}{\angstrom}               \\
1GZM  &\num{349}   &\num{7}     & \SI{62}{\percent}        & X-ray & \SI{2.7}{\angstrom}               \\
1J4N  &\num{116}   &\num{4}     & \SI{80}{\percent}        & X-ray & \SI{2.2}{\angstrom}               \\
1KPL  &\num{203}   &\num{8}     & \SI{76}{\percent}        & X-ray & \SI{3.0}{\angstrom}               \\
1OCC  &\num{191}   &\num{5}     & \SI{74}{\percent}        & X-ray & \SI{2.8}{\angstrom}               \\
1OKC  &\num{297}   &\num{9}     & \SI{71}{\percent}        & X-ray & \SI{2.2}{\angstrom}               \\
1PV6  &\num{189}   &\num{8}     & \SI{87}{\percent}        & X-ray & \SI{3.5}{\angstrom}               \\
1PY6  &\num{227}   &\num{9}     & \SI{75}{\percent}        & X-ray & \SI{1.8}{\angstrom}               \\
1RHZ  &\num{166}   &\num{5}     & \SI{65}{\percent}        & X-ray & \SI{3.5}{\angstrom}               \\
1U19  &\num{278}   &\num{7}     & \SI{66}{\percent}        & X-ray & \SI{2.2}{\angstrom}               \\
1XME  &\num{568}   &\num{18}    & \SI{79}{\percent}        & X-ray & \SI{2.3}{\angstrom}               \\
2BG9  &\num{91}    &\num{3}     & \SI{87}{\percent}        & EM    & ---                  \\
2BL2  &\num{145}   &\num{4}     & \SI{88}{\percent}        & X-ray & \SI{2.1}{\angstrom}               \\
2BS2  &\num{217}   &\num{8}     & \SI{80}{\percent}        & X-ray & \SI{1.8}{\angstrom}               \\
2IC8  &\num{182}   &\num{7}     & \SI{68}{\percent}        & X-ray & \SI{2.1}{\angstrom}               \\
2K73  &\num{164}   &\num{5}     & \SI{62}{\percent}        & NMR   & ---                  \\
2KSF  &\num{107}   &\num{4}     & \SI{64}{\percent}        & NMR   & ---                  \\
2KSY  &\num{223}   &\num{7}     & \SI{78}{\percent}        & NMR   & ---                  \\
2NR9  &\num{196}   &\num{8}     & \SI{75}{\percent}        & X-ray & \SI{2.2}{\angstrom}               \\
2XUT  &\num{524}   &\num{16}    & \SI{72}{\percent}        & X-ray & \SI{3.6}{\angstrom}               \\
3GIA  &\num{433}   &\num{15}    & \SI{81}{\percent}        & X-ray & \SI{2.2}{\angstrom}               \\
3KCU  &\num{285}   &\num{10}    & \SI{67}{\percent}        & X-ray & \SI{2.2}{\angstrom}               \\
3KJ6  &\num{366}   &\num{8}     & \SI{47}{\percent}        & X-ray & \SI{3.4}{\angstrom}               \\
3P5N  &\num{189}   &\num{6}     & \SI{70}{\percent}        & X-ray & \SI{3.6}{\angstrom}               \\
\midrule
2BHW  &\num{669}   &\num{12}    & \SI{45}{\percent}        & X-ray & \SI{2.5}{\angstrom}               \\
2H8A  &\num{363}   &\num{12}    & \SI{79}{\percent}        & EM    & \SI{3.2}{\angstrom}               \\
2HAC  &\num{66}    &\num{2}     & \SI{79}{\percent}        & NMR   & ---                  \\
2L35  &\num{95}    &\num{3}     & \SI{81}{\percent}        & NMR   & ---                  \\
2ZY9  &\num{344}   &\num{16}    & \SI{90}{\percent}        & X-ray & \SI{2.9}{\angstrom}               \\
3CAP  &\num{696}   &\num{18}    & \SI{68}{\percent}        & NMR   & \SI{2.9}{\angstrom}               \\
\bottomrule
\end{tabular}

  \caption[Proteins used for benchmarking the \gls{epr} BCL::MP-Fold algorithm]{\textbf{Proteins
      used for benchmarking the structure prediction algorithm.} The twenty-nine proteins for the
    benchmark were chosen to cover a wide range of sequence length, number of \glspl{sse} as well
    as number and percentage of residues within \glspl{sse} while having a mutual sequence identity
    of less than \SI{20}{\percent}. The columns denote the sequence length, the number of
    \glspl{sse}, the number of residues within \glspl{sse}, and the percentage of the residues is
    within \glspl{sse}. The proteins above the separating line are monomeric proteins; below the
    separating line are multimeric proteins. 2HAC, 2ZY9, and 3CAP are homodimers, 2BHW and 2H8A are
    homotrimers, and 2L35 is a heterodimer. 1GZM was additionally included to evaluate the protocol
    on experimentally determined data.}
  \label{tab:mpepr_benchmark_set}
\end{wraptable}

One critical limitation for \emph{de novo} protein structure prediction from \gls{epr} data is that
measurements relate to the tip of the spin label side-chain where the unpaired electron is located
whereas information of the placement of backbone atoms is needed to define the protein fold. For
distance measurements, this introduces an uncertainty in relating the distance measured between the
two spin labels to a distance between points in the backbone of the protein. This uncertainty,
defined as the difference between the distance between the spin labels and the distance between the
corresponding $\mathrm{C_\beta}$-atoms is up to \SI{12}{\angstrom} \autocite{Alexander2008,
  Hirst2011}. To address this uncertainty we previously introduced a \gls{cone} model, which
provides a knowledge-based probability distribution for the $\mathrm{C_\beta}$-atom distance given
an \gls{epr}-measured spin label distance \autocite{Alexander2008, Vera2013}. Using the \gls{cone}
model, just twenty-five or even eight \gls{epr} measured distances for T4-lysozyme, enabled Rosetta
to provide models matching the experimentally determined structure to atomic detail including
backbone and side-chain placement \autocite{Alexander2008}. Further success was reported by Yang
\emph{et al.}~\autocite{Yang2010}, who successfully determined the tertiary structure of a
homodimer by using inter-chain restraints determined from \gls{nmr} and \gls{epr}
experiments. These studies demonstrate that \emph{de novo} prediction methods can supplement
\gls{epr} data sufficiently to allow structure elucidation of a protein.

\emph{De novo} membrane protein structure prediction was demonstrated with Rosetta using twelve
proteins with multiple transmembrane spanning helices \autocite{Yarov-Yarovoy2006}. The method was
generally successful for the membrane topology for small proteins up to \num{278} residues. The
results of the study suggest that sampling of large membrane topologies requires methods that
directly sample structural contacts between sequence distance regions of the protein
\autocite{Barth2009}.

For this purpose, we developed an algorithm that assembles protein topologies from \glspl{sse}
termed BCL::Fold \autocite{Karakas2012}. The omission of loop regions in the initial protein
folding simulation allows sampling of structural contacts between regions distant in sequence and
thereby rapidly enumerates all likely protein topologies. A knowledge-based potential guides the
algorithm towards physically realistic topologies. The algorithm is particularly applicable for the
determination of membrane protein topologies as transmembrane spans are dominated by regularly
ordered \glspl{sse} \autocite{Weiner2013}. Loop regions and amino acid side-chains can be added in
later stages of modeling structure. The algorithm was tested in conjunction with medium-resolution
density maps \autocite{Lindert2009a} achieving models accurate at atomic detail in favorable cases
\autocite{Lindert2012a}. The algorithm was also tested in conjunction with sparse \gls{nmr} data
\autocite{Weiner2014}.

The present study combines \gls{epr} distance and accessibility restraints with the BCL::Fold
\gls{sse} assembly methodology for the prediction of membrane protein topologies. In a first step,
we introduce scores specific to \gls{epr} distances and accessibilities and demonstrate their
ability to enrich for accurate models. In a second step, we describe the approach and results for
assembling twenty-three monomeric and six multimeric membrane proteins guided by \gls{epr} distance
and accessibility restraints. The results demonstrate that the inclusion of protein specific
structural information improves the frequency with which accurate models are sampled and greatly
improves the discrimination of incorrect models.

\section{Materials and methods}

\subsection{Compilation of the benchmark set}

Twenty-nine membrane proteins of known structure were used to demonstrate the ability of \gls{epr}
specific scores to improve sampling during protein structure prediction as well as selecting the
most accurate models. The proteins for the benchmark were chosen to cover a wide range of sequence
length, number of \glspl{sse}, and percentage of residues within \glspl{sse}
(\Fref{tab:mpepr_benchmark_set}). Twenty-three of the proteins were monomers ranging in size from
\numrange{91}{568} residues. One protein (2L35) has two chains, with the second chain being a
single transmembrane span. The remaining five proteins were symmetric multimeric proteins of two or
three subunits containing up to \num{696} residues. \num{5000} independent structure prediction
trajectories were conducted for each protein without restraints, with distance restraints only,
with accessibility restraints only, and with distance and accessibility restraints. In order to
achieve results that are independent of one specific spin labelling pattern, ten different
restraint sets were used for each protein. Those trajectories were conducted with \glspl{sse}
predicted from sequence and, to test the influence of incorrectly predicted secondary structure,
with the \glspl{sse} obtained from the experimentally determined structure. In addition, rhodopsin
(\Gls{pdb} entry 1GZM) was added to the benchmark set to demonstrate the algorithm’s ability to
work with experimentally determined restraints.

\subsection{Simulation of \gls{epr} restraints}

For 1GZM, \gls{epr} distance restraints were available \autocite{Altenbach2008}, whereas for the
other proteins \gls{epr} distance and accessibility restraints were simulated to obtain data sets
for each of the twenty-nine proteins. Accessibility restraints were simulated by calculating the
neighbor vector value \autocite{Durham2009} for residues within \glspl{sse} of each protein. Unlike
the neighbor count approximation of the \gls{sasa}, the neighbor vector approach takes the relative
placement of the neighbors with respect to the vector from the $\mathrm{C_\alpha}$-atom to the
$\mathrm{C_\beta}$-atom into account. It thereby becomes a more accurate predictor of \gls{sasa}
\autocite{Durham2009}. The resulting exposure value for each residue was considered an oxygen
accessibility measurement. One restraint per two residues within the transmembrane segment of each
\gls{sse} was simulated.

Distance restraints were simulated using a restraint selection algorithm \autocite{Kazmier2011},
which distributes measurements across all \glspl{sse} (\Fref{lst:mpepr_dataset_generation}). It
also favors measurements between residues that are far apart in sequence. One restraint was
generated per five residues within the transmembrane segment of an \gls{sse}, if not indicated
otherwise. Distances are calculated between the $\mathrm{C_\beta}$-atoms; for glycine, the
$\mathrm{H_{\alpha2}}$-atom is used. To simulate a likely distance observed in an actual \gls{epr}
experiment, the distance is adjusted by an amount selected randomly from the probability
distribution of observing a given difference between the spin-spin distance ($D_{\mathit{SL}}$) and
the back bone distance ($D_{\mathit{BB}}$) \autocite{Hirst2011}. In order to reduce the possibility
of bias arising from restraint selection and spin labelling patterns, ten independent restraint
sets were generated. For the five symmetric multimeric proteins, the same protocol was used, but
only distance restraints between the same residues in the different subunits were considered.

\subsection{Translating \gls{epr} accessibilities into structural restraints}

\begin{wrapfigure}{R}{0.55\textwidth}
  \centering
  \includegraphics[width=0.53\textwidth]{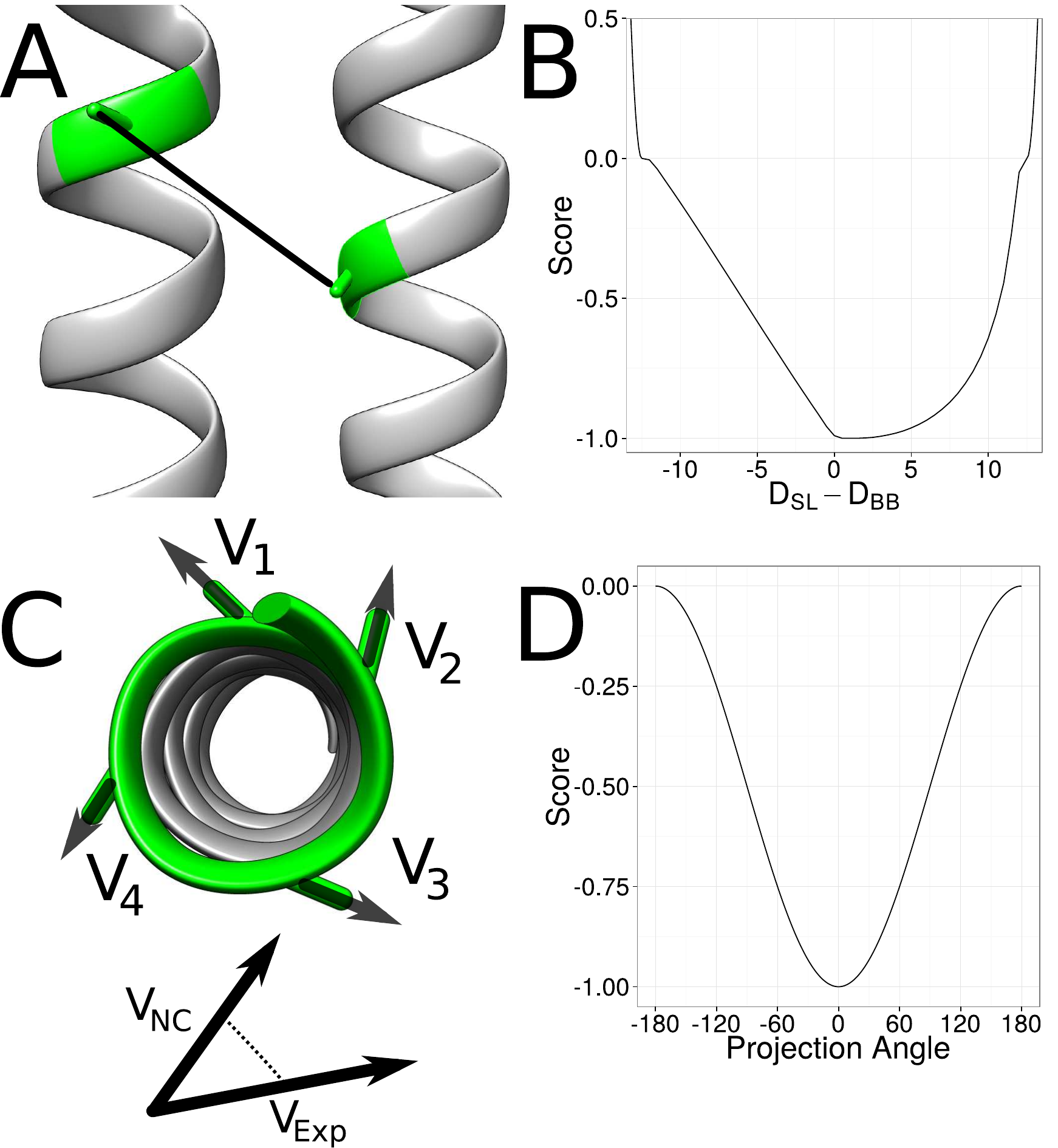}
  \caption[Translation from EPR data into structural restraints]{\textbf{Translation from \gls{epr}
      data into structural restraints.} \Gls{epr} distance measurements measure distances between
    residues in a protein indirectly. Whereas the experiment determined the spin-spin distance
    ($D_{\mathit{SL}}$), a distance between the backbone atoms ($D_{\mathit{BB}}$) is needed during
    the \emph{de novo} protein structure prediction process. Therefore a translation from
    $D_{\mathit{SL}}$ to $D_{\mathit{BB}}$ is necessary. BCL::Fold uses a knowledge-based potential
    to evaluate the agreement of the distance between the $\mathrm{C_\beta}$-atoms in the model
    with the experimentally determined spin-spin distances (B). \Gls{epr} accessibility data is
    translated into structural restraints by summing up the hydrophobic moment vectors
    ($\mathrm{C_\alpha}$-atom to $\mathrm{C_\beta}$-atom) of four consequtive residues (C). This is
    done twice: first the normalized $\mathrm{C_\alpha}$-$\mathrm{C_\beta}$ vectors are multiplied
    with the accessibility determined in the \gls{epr} experiment, the second time they are
    multiplied with the neighbor count of the residue in the model. The vectors are summed up for
    each approach and the projection angle between the two resulting vectors is scored, with an
    angle of \SI{0}{\degree} being the best and \SI{180}{\degree} being the worst agreement (D).}
  \label{fig:mpepr_translation_to_score}
\end{wrapfigure}

\Gls{epr} accessibility measurements are typically made in a sequence scanning fashion over a
portion of the target protein. Although each individual accessibility measurement is difficult to
interpret, the pattern of accessibilities over a stretch of amino acids within an \gls{sse}
indicates reliably, which phase of the \gls{sse} is exposed to solvent/membrane versus buried in
the protein core. We found accessibility restraints to have a limited impact on structure
prediction for soluble proteins \autocite{Alexander2008}. We concluded that this is the case as
knowledge-based potentials on their own can distinguish the polar phase of an \gls{sse} that is
exposed to an aqueous solvent from a hydrophobic phase buried in the protein core. However, we also
hypothesized that the situation will be different for membrane proteins where it would be harder to
distinguish the membrane-exposed from the buried phase of an $\mathrm{\alpha}$-helix as both of
these tend to be apolar.

Our approach for developing an \gls{epr} accessibility score takes advantage of the regular
geometry within the \gls{sse}: The exposure moment of a window of amino acids is defined as
$E_w = \sum\nolimits_{n=1}^N e_n \times s_n$, where $N$ is the number of residues in the window,
$e_n$ is the exposure value of residue $n$, and $s_n$ is the normalized vector from the
$\mathrm{C_\alpha}$-atom to the $\mathrm{C_\beta}$-atom of residue $n$. This equation was inspired
by the hydrophobic moment as previously defined \autocite{Eisenberg1984}. The exposure moment
calculated from solvent accessible surface area \gls{sasa} has been previously demonstrated to
approximate the moment calculated from \gls{epr} accessibility measurements
\autocite{Salwinski1999}.

During \emph{de novo} protein structure prediction, the protein is represented only by its backbone
atoms hampering calculation of \gls{sasa}. Further, calculation of \gls{sasa} from an atomic detail
model would be computationally prohibitive for a rapid scoring function in \emph{de novo} protein
structure prediction. Therefore, the neighbor vector approximation for \gls{sasa} is used
\autocite{Durham2009}. The exposure moment is calculated for overlapping windows of length seven
for $\mathrm{\alpha}$-helices and four for $\mathrm{\beta}$-strands. The score is computed as
$S_{orient} = -0.5 \times cos(\theta)$ where $\theta$ is the torsion angle between the exposure
moments. This procedure assigns a score of \num{-1} if $\theta = \SI{0}{\degree}$ and a score of
\num{0} if $\theta = \SI{180}{\degree}$ (\Fref{fig:mpepr_translation_to_score}).

It has previously been demonstrated that the burial of sequence segments relative to other segments
can be determined from the average accessibility values measured for that stretch of sequence
\autocite{Chakrapani2008}. To capture this information, the magnitude of the exposure moment for
overlapping residue windows is determined from the model structure and from the measured
accessibility. The Pearson correlation is then calculated between the rank order magnitudes of the
structural versus experimental moments. This gives a value between \num{-1}, which indicates the
structural and exposure magnitudes are oppositely ordered, and \num{1}, which means the structural
and exposure magnitudes are ordered equivalently. The score $\mathrm{S_{\mathit{magn}}}$ is
obtained by negating the resulting Pearson correlation value so that matching ordering will get a
negative score and be considered favorable.

\subsection{Translating \gls{epr} distances into structural restraints}

The \gls{cone} model \autocite{Alexander2008} yields a predicted distribution for the difference
between $D_{\mathit{SL}}$ and $D_{\mathit{BB}}$. This distribution was converted into a
knowledge-based potential function, which is used to score the agreement of models with
experimentally determined \gls{epr} distance restraints \autocite{Hirst2011}. This score spans a
range of $D_{\mathit{SL}} - D_{\mathit{BB}}$ between \SI{-12}{\angstrom} and
\SI{+12}{\angstrom}. $D_{\mathit{SL}}$ is the \gls{epr} measured distance between the two spin
labels; $D_{\mathit{BB}}$ is the distance between the corresponding $\mathrm{C_\beta}$- or
$\mathrm{H_{\alpha2}}$-atoms on the residues of interest; $D_{\mathit{SL}}-D_{\mathit{BB}}$ is the
difference between these two distances (\Fref{fig:mpepr_translation_to_score}).

In addition, we found it beneficial to add an attractive potential on either side of the range
spanned by the scoring function to provide an incentive for the \gls{mcm} minimization to bring
structures within the defined range of the scoring function. These attractive potentials use a
cosine function to transition between a most unfavorable score of \num{0} and a most favorable
score of \num{-1}. The attractive potential is positive for
$\SI{30}{\angstrom} \geq \mid D_{\mathit{SL}} - D_{\mathit{BB}} \mid \geq \SI{12}{\angstrom}$. It
levels to \num{0} when the difference between $D_{\mathit{BB}}$ and $D_{\mathit{SL}}$ approaches
\SI{12}{\angstrom} (\Fref{fig:mpepr_translation_to_score}).

\subsection{Summary of the folding protocol} \label{sec:mpepr_summary_protocol}

\begin{wrapfigure}{R}{0.5\textwidth}
  \centering
  \includegraphics[width=0.48\textwidth]{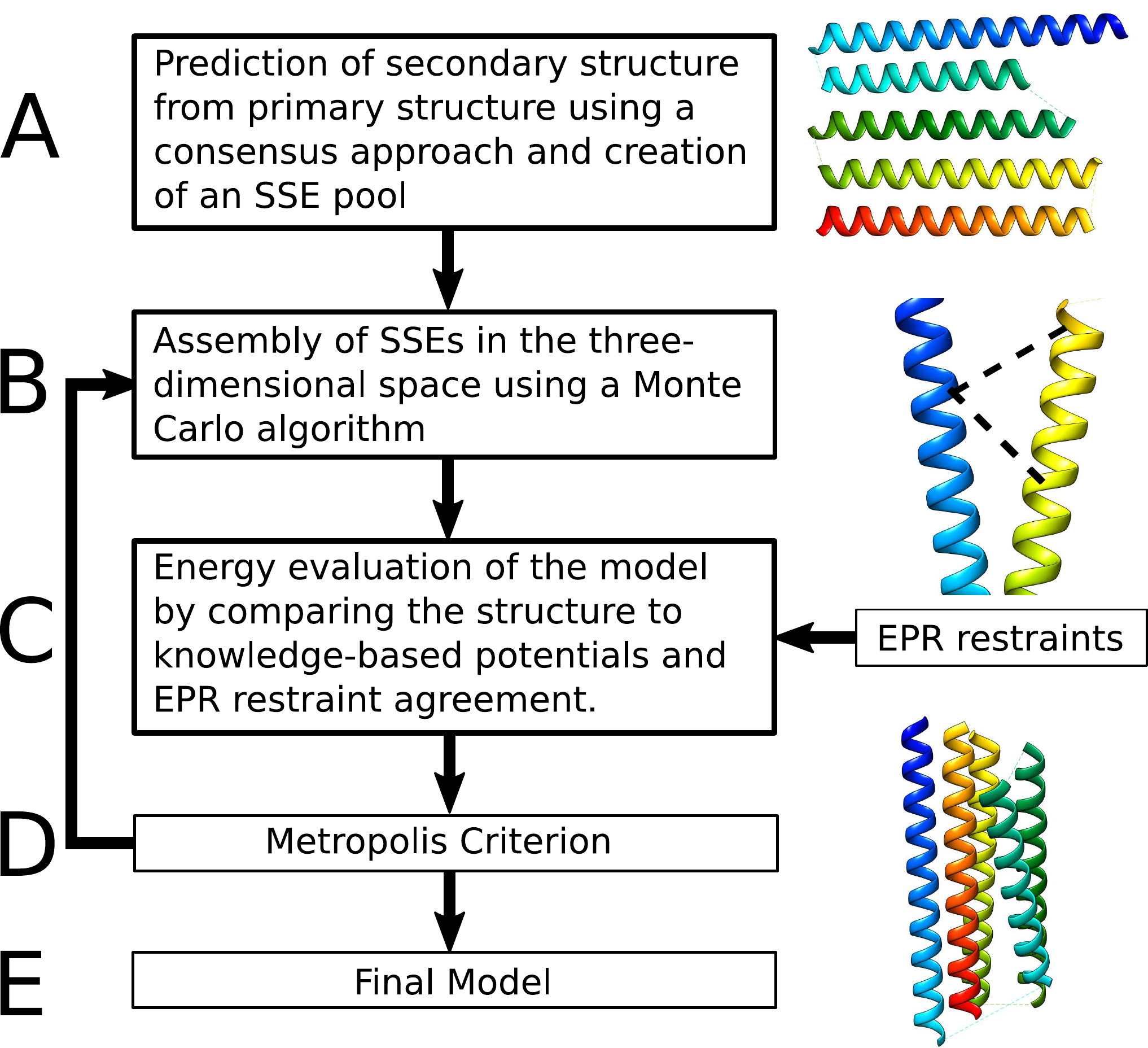}
  \caption[Structure prediction protocol for using \gls{epr} data]{\textbf{Structure prediction
      protocol for using \gls{epr} data.} BCL::Fold assembles predicted \glspl{sse} in the
    three-dimensional space to predict the tertiary structure of a protein. In a first step, the
    secondary structure is predicted using a consensus of several \gls{sse} prediction methods like
    \gls{octopus} and Jufo9D (A). Consequently, the predicted \glspl{sse} are added to the model
    and transformed using an \gls{mcm} algorithm (B). The outcome of each transformation is
    evaluated with knowledge-based potential functions scoring \gls{sse} packing, radius of
    gyration, amino acid exposure, and amino acid pairing, loop closure geometry, secondary
    structure length and content, \gls{sse} clashes, and the agreement of the model with the
    provided \gls{epr} distance and accessibility restraints (C). Based on the difference in score
    between the model before and the model after applying the transformation the outcome is either
    accepted or rejected (D). This process is repeated until a specified number of iterations or a
    maximum number of steps without score improvement is reached. The resulting models are then
    ranked based on their score according to the knowledge-based potential functions (E).}
  \label{fig:mpepr_protocol}
\end{wrapfigure}

The protein structure prediction protocol (\Fref{fig:mpepr_protocol}) is based on the protocol of
BCL::Fold for soluble proteins \autocite{Karakas2012}. The method assembles \glspl{sse} in the
three-dimensional space, drawing from a pool of predicted \glspl{sse}. A \gls{mc} energy
minimization with the Metropolis criteria is used to search for models with favorable
energies. Models are scored after each \gls{mc} step using knowledge-based potentials describing
optimal \gls{sse} packing, radius of gyration, amino acid exposure, and amino acid pairing, loop
closure geometry, secondary structure length and content, and penalties for clashes
\autocite{Wotzel2012}.

The algorithm was adapted for membrane protein folding by altering the amino acid exposure
potential according to an implicit membrane environment \autocite{Weiner2013}. Additional scores
are used, which favor orthogonal placement of \glspl{sse} relative to the membrane
($\mathrm{\mathit{SSE}_\mathit{align}}$) and penalizing models with loops going through the
membrane ($\mathrm{\mathit{MP}_\mathit{top}}$). All moves introduced for soluble proteins are used
\autocite{Karakas2012}. In addition, we include perturbations that optimize the placement of the
protein in the membrane such as translation of individual \glspl{sse} in the membrane as well as
rigid body translation and rotation of the entire protein.

The assembly of the protein structure is broken down into five stages of sampling with large
structural perturbation moves that can alter the topology of the protein. Each of the five stages
lasts for a maximum of \num{2000} \gls{mc} steps. If an energetically improved structure has not
been generated within the previous \num{400} \gls{mc} steps, the minimization for that stage will
cease. Over the course of the five assembly stages, the weight of clashing penalties in the total
score is ramped as \numlist{0;125;250;375;500}.

Following the five stages of protein assembly, a structural refinement stage takes place. This
stage lasts for a maximum of \num{2000} \gls{mc} steps and will terminate sooner if an
energetically improved model is not sampled within the previous \num{400} steps. The refinement
stage consists of small structural perturbations, which will not drastically alter the topology of
the protein model.

After \num{5000} models have been generated for each protein, the models are filtered according to
\gls{epr} distance score. The top \SI{10}{\percent} or \num{500} models resulting from the
structure prediction protocol are selected for a second round of energy minimization. The second
round occurs as described above, the only difference being that the minimization uses the \gls{sse}
placements of a given protein as a starting point. For each starting structure, \num{10} models are
created, resulting in \num{5000} models. This boot strapping approach, which re-optimizes
structures that are in good agreement with the \gls{epr} restraints and with the knowledge-based
potential was beneficial when combining BCL::MP-Fold with limited \gls{nmr} data and is not applied
when no experimental data are used \autocite{Weiner2014}.

\subsection{Summary of the benchmark setup}

To test the influence of \gls{epr} restraints, each protein besides 1GZM was folded in the absence
of restraints, with just distance restraints, with just accessibility restraints, and with distance
and accessibility restraints. To test the influence of secondary structure prediction accuracy (see
\fref{sec:mpepr_protocol}), the experiment was repeated with optimal \glspl{sse} derived from the
experimentally determined structure. 1GZM was only folded without restraints and with the
experimentally determined distance restraints. \num{5000} models were created for each of the
benchmark proteins in independent \gls{mcm} folding trajectories. \Gls{epr} distance and
accessibility scores are used during the five assembly and one refinement stages of structure
prediction protocol. The \gls{epr} distance scores have a weight of \num{40} during all assembly
and refinement stages using either pool.

\subsection{Structure prediction protocol} \label{sec:mpepr_protocol}

For each protein, two sets of \gls{sse} pools are generated for use during structure assembly. The
first \gls{sse} pool consists of the transmembrane spanning helices as predicted by
\gls{octopus}. The second \gls{sse} pool contains elements predicted by \gls{octopus} as well as
\glspl{sse} predicted from sequence by Jufo9D (\Fref{lst:mpepr_sse_generation}). Using these two
\gls{sse} pools, the structure prediction protocol is independently conducted twice:
\begin{inparaenum}[a)]
\item once using the \gls{sse} pool containing predictions from \gls{octopus} and Jufo9D (``full
  pool'') and
\item once emphasizing the predictions by \gls{octopus} (``OCTOPUS pool'').
\end{inparaenum}
Emphasis is placed on \gls{octopus} predictions by using only the \gls{octopus} generated \gls{sse}
pool during the first two stages of assembly. During last three stages of structure assembly, the
\glspl{sse} predicted from Jufo9D are added to the pool. This allows for better coverage of
\glspl{sse} within the structure, since \gls{octopus} only predicts transmembrane spanning helices.

\Gls{epr} specific scores are used during the five assembly and one refinement stages of structure
prediction (\Fref{lst:mpepr_structure_prediction}). The \gls{epr} distance scores have a weight of
\num{40} over the course of the assembly and refinement stages.

\subsection{Calculating \gls{epr} score enrichments}

The enrichment value is used to evaluate how well a scoring function is able to select the most
accurate models from a given set of models. The models of a given set are sorted by their
\gls{rmsd100} values. The \SI{10}{\percent} of the models with the lowest \gls{rmsd100} values put
into the set $P$ (positive) the rest of the models will be put into the set $N$ (negative). The
models of $S$ are then also sorted by their assigned scoring value and the \SI{10}{\percent} of the
models with the lowest (most favorable) score are put into the set $T$. The models, which are in
$P$ and in $T$ are the models, which are correctly selected by the scoring function and their
number will be referred to as $\mathit{TP}$ (true positives). The number of models, which are in
$P$ but not in $T$ are the models, which are not selected by scoring function despite being among
the most accurate ones. They will be referred to as $\mathit{FN}$ (false negative). The enrichment
will then be calculated as $e = \frac{\#\mathit{TP}}{\#P} \times \frac{\#P+\#N}{\#P}$. The positive
models are in this case considered the \SI{10}{\percent} of the models with the lowest
\gls{rmsd100} values. Therefore, $\frac{\#P+\#N}{\#P}$ is a constant value of \num{10.0}. No
enrichment would be a value of \num{1.0} and an enrichment value between \num{0.0} and \num{1.0}
indicates that the score selects against accurate models.

\section{Results}

\subsection{Using \gls{epr} specific scores during membrane protein structure prediction improves
  sampling accuracy}

For each protein, the ten models sampled with the best \gls{rmsd100} \autocite{Carugo2001} values
are used to determine ability to sample accurate models by taking their \gls{rmsd100} value
average, $\mu_{10}$. Using the best ten models by \gls{rmsd100} provides a more consistent measure
of sampling accuracy compared to looking at the single best because of the random nature of the
structure prediction protocol. Additionally, the percentages of models with an \gls{rmsd100} less
than \SI{4}{\angstrom} and less than \SI{8}{\angstrom}, $\tau_4$ and $\tau_8$, were calculated.

\begin{figure}
  \centering
  \includegraphics[width=\linewidth]{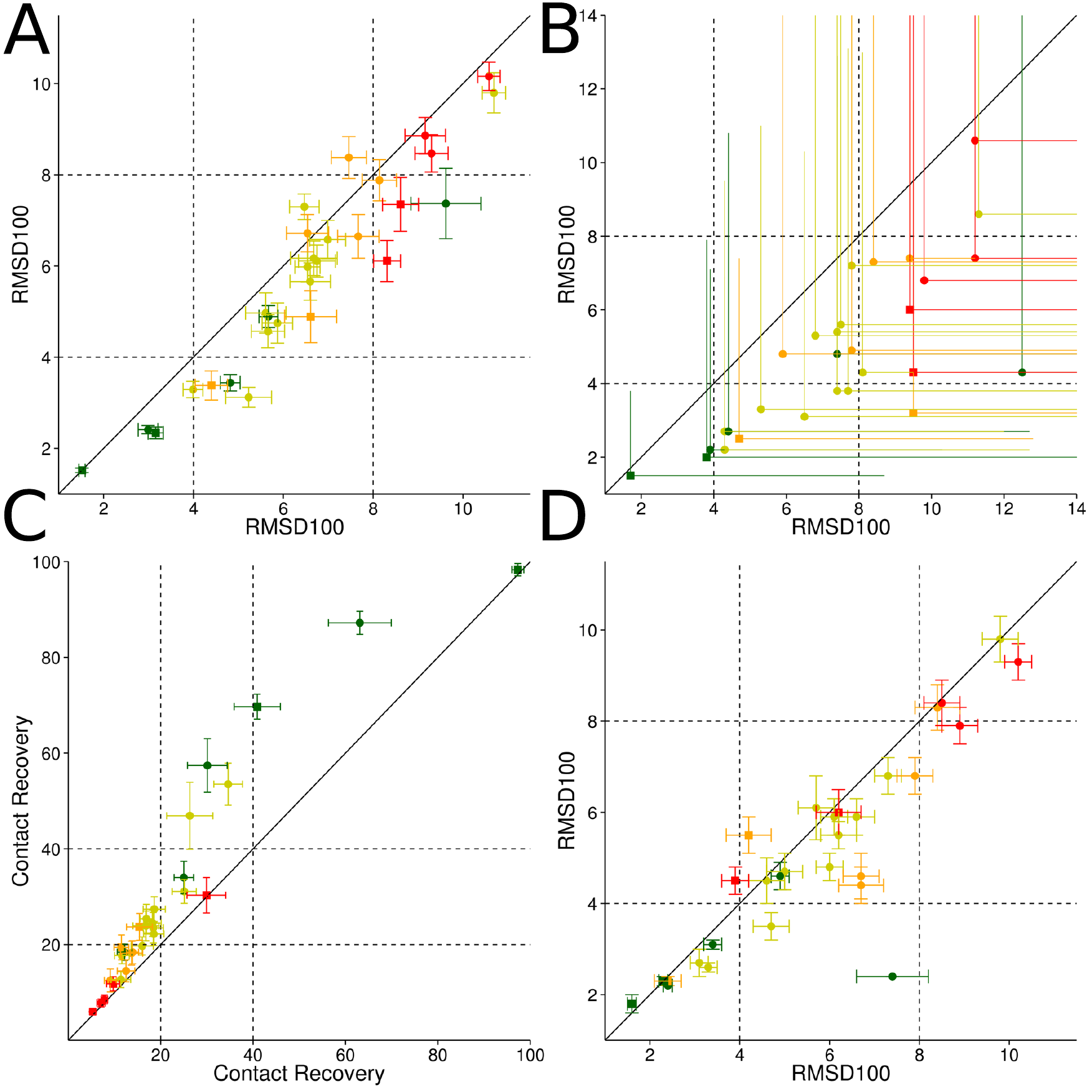}
  \caption[Sampling accuracy, contact recovery, and enrichment results when using EPR
  data]{\textbf{Sampling accuracy, contact recovery, and enrichment results when using \gls{epr}
      data.} By using \gls{epr} distance and accessibility data in the structure prediction process
    the sampling accuracy can be improved significantly for monomeric (circles) as well as
    oligomeric (squares) proteins (A). The sampling accuracy could be improved in twenty-five out
    of twenty-nine cases by using \gls{epr} distance and accessibility data, which is demonstrated
    by comparing the average \gls{rmsd100} values of the \SI{1}{\percent} most accurate models
    predicted without (x-axis) and with \gls{epr} data (y-axis) in (A). Adding protein specific
    structural information in the form of \gls{epr} distance and accessibility restraints also
    improves our ability to select the most accurate models among the sample ones. In each of the
    twenty-nine cases \gls{epr} distance and accessibility restraints enable us to select more
    accurate models when compared to structure prediction without \gls{epr} data available. Shown
    are the average (line) and best (dot/square) \gls{rmsd100} values of the best \SI{1}{\percent}
    models by BCL score with (y-axis) and without (x-axis) \gls{epr} restraints (B). By using
    \gls{epr} accessibility data only (y-axis) the Contact Recovery could be improved in twenty-two
    out of twenty-nine cases (C) when compared to structure prediction without \gls{epr}
    accessibility restraints (x-axis). Improvements in \gls{sse} prediction methods would also lead
    to improved sampling accuracies (D, see also \fref{tab:mpepr_contact_recovery_native}). In
    twenty-one out of twenty-nine cases the average \gls{rmsd100} of the ten most accurate models
    could be improved by using \gls{sse} definitions obtained from the experimentally determined
    structure (y-axis) compared to using predicted \glspl{sse}.}
  \label{fig:mpepr_results}
\end{figure}

By using \gls{epr} distance and accessibility scores, not only is the frequency increased with
which higher accuracy models are sampled, but the best models achieve an accuracy not sampled in
the absence of \gls{epr} data (\Fref{tab:mpepr_sampling}). Across all proteins, $\mu_{10}$ is, on
average, \SI{6.0}{\angstrom} when \gls{epr} distance and accessibility scores are not used. When
adding restraints for distances and then both distances and accessibilities, the average $\mu_{10}$
value drops to \SIlist{5.1;5.0}{\angstrom}, respectively (\Fref{tab:mpepr_sampling}). By only
adding \gls{epr} accessibility restraints the average $\mu_{10}$ over all proteins improves only
slightly to \SI{5.8}{\angstrom}. This demonstrates that the accuracy of the models is primarily
improved by using \gls{epr} distance restraints in the structure prediction process. With the
exception of 1KPL and 2XUT, all proteins achieve a $\mu_{10}$ value of less than
\SI{8.0}{\angstrom}. This indicates the placement of the transmembrane spanning regions follow the
experimentally determined structures and the correct fold could be
predicted. \Fref{fig:mpepr_results} compares the \gls{rmsd100} values of the average of the
\SI{1}{\percent} most accurate models with and without the usage of \gls{epr} distance restraints
--- an average improvement of \SI{0.8}{\angstrom} over the benchmark set is observed. The shift to
lower \gls{rmsd100} values in distributions for selected benchmark proteins is shown in
\fref{fig:mpepr_results}. The average $\tau_4$ and $\tau_8$ values improve from
\SIlist{3;13}{\percent}, when folding without \gls{epr} restraints, and to \SIlist{6;19}{\percent}
when using \gls{epr} restraints, respectively.

The six multimeric proteins achieve an average $\mu_{10}$ value of \SI{5.0}{\angstrom} when the
structure prediction was conducted without using \gls{epr} restraints. By using \gls{epr} distance
and accessibility restraints $\mu_{10}$ could be improved to \SI{2.9}{\angstrom}. The $\tau_4$ and
$\tau_8$ values could be improved from \SIlist{13;24}{\percent} to \SIlist{21;41}{\percent} when
using \gls{epr} distance and accessibility restraints in the structure prediction process.

\subsection{\Gls{epr} accessibility scores are important for improving contact recovery}

\Gls{epr} accessibility scores were previously used in conjunction with the Rosetta protein
structure prediction algorithm \autocite{Alexander2008}. The scores were applied in a benchmark to
predict the structures of the small soluble proteins T4-lysozyme and
$\mathrm{\alpha}$A-crystallin. The improvement in sampling models that are more accurate was
compared between prediction trajectories using an \gls{epr} distance score and trajectories using
an \gls{epr} distance score coupled with an accessibility score. For T4-lysozyme and
$\mathrm{\alpha}$A-crystallin, using the accessibility score did not result in a significant
improvement in the accuracy of models sampled. This was attributed to the simple rule of exposure
that is well captured by the knowledge-based potentials: polar residues tend to be exposed to
solvent; apolar residues tend to be buried in the core of the protein.

Membrane proteins are subjected to a more complex set of possible environments. Any given residue
can reside buried in the core of the protein or exposed to different environments ranging from the
membrane center to a transition region to an aqueous solvent. If the protein fold contains a pore,
a residue can be solvent-exposed deep in the membrane \autocite{Dalmas2010}. Such a complex
interplay of environments will not be as easily distinguished by knowledge-based potentials. Here
it has been demonstrated that using \gls{epr} accessibility information consistently improves the contact
recovery for highest accurate models.

Although improvements regarding sampling accuracy and selection of the most accurate models by
\gls{rmsd100} is mainly achieved by using \gls{epr} distance restraints, \gls{epr} accessibility
restraints help determining the correct rotation state of \glspl{sse} and therefore improves the
number of recovered contacts (\Fref{fig:mpepr_results}). A contact is defined as being between
amino acids, which are separated by at least six residues and have a maximum Euclidean distance of
\SI{8}{\angstrom}. We are measuring the percentage of the contacts in the experimentally determined
protein structure, which could be recovered in the models. In order to be independent of huge
deviations occurring when only looking at the best model sampled, we quantify the average contact
recovery of the ten models with the highest contact recovery ($\phi_{10}$) and the percentage of
models, which have more than \SIlist{20;40}{\percent} of the contacts recovered ($\gamma_{20}$ and
$\gamma_{40}$).

For folding without \gls{epr} restraints, the average $\phi_{10}$ value over all twenty-three
monomeric proteins was \SI{23}{\percent} whereas with accessibility restraints it was
\SI{31}{\percent} (\Fref{tab:mpepr_contact_recovery}). Using distance restraints additionally to
the accessibility restraints $\phi_{10}$ remains at \SI{31}{\percent}. This is demonstrating that
improvements in contact recovery are mainly achieved by using \gls{epr} accessibility restraints in
the structure prediction process. The average $\gamma_{20}$ and $\gamma_{40}$ values over all
twenty-nine proteins for structure prediction without \gls{epr} restraints were
\SIlist{5;3}{\percent}. By using \gls{epr} accessibility restraints, the values could be improved
to \SIlist{12;16}{\percent}, respectively.

For the six multimeric proteins, improvements in contact recovery by the usage of \gls{epr}
accessibility restraints are observed as $\phi_{10}$, $\gamma_{20}$, and $\gamma_{40}$ values could
be increased to \SIlist{46;25;16}{\percent} from the previous values of \SIlist{38;17;14}{\percent}
when performing protein structure prediction without \gls{epr} data. By complementing the
accessibility with distance restraints $\phi_{10}$, $\gamma_{20}$, and $\gamma_{40}$ values can be
improved to \SIlist{50;30;16}{\percent}.

\subsection{\Gls{epr} specific scores select for accurate models of membrane proteins}

The ability of \gls{epr} specific scores to select for accurate models is tested by calculating
enrichment values for structure prediction trials of twenty-nine membrane proteins
(\Fref{tab:mpepr_enrichment}). The enrichment of a scoring function indicates how well the score
identifies a protein model that is accurate by a good score. It computed as the cardinality of the
intersection $I=\mathit{HS} \cap P$ with $P$ being the set of the accurate models and $\mathit{HS}$
being the set of the \SI{10}{\percent} of the models with the most favorable score (see
\fref{sec:mpepr_summary_protocol}) \autocite{Wotzel2012}. Accurate is defined as the
\SI{10}{\percent} of the models with the lowest \gls{rmsd100} when compared to the experimentally
determined structure. Therefore, if a score correctly identifies all accurate models as being
accurate, a perfect enrichment would result in a value of \num{10.0}.

Enrichment values are computed for the protein models created without experimental restraints. For
protein structure prediction without \gls{epr} data, the average enrichment value for just the
knowledge-based potentials over all twenty-nine proteins is \num{1.3}. By using \gls{epr} distance
and accessibility data, the average enrichment is improved to \num{2.5}. The enrichment for using
\gls{epr} distance and accessibility restraints ranges from \numrange{1.1}{6.2}. In seventeen out
of twenty-nine cases, the enrichment is greater than \num{2.0}. In twenty-three out of twenty-nine
cases the enrichment could be improved by at least \num{0.5} (\Fref{tab:mpepr_enrichment}). By
using \gls{epr} accessibility data only the average enrichment over all proteins is \num{1.6},
demonstrating that improvements regarding the selection of the most accurate models are mainly
caused by \gls{epr} distance restraints.

\subsection{The number of restraints determines the significance of improvements in sampling
  accuracy}

\begin{table}
  \centering
  \small
  \begin{tabular}{lrrrrrrrrr}
\toprule
  & \multicolumn{3}{c}{1/10} & \multicolumn{3}{c}{1/3} & \multicolumn{3}{c}{1/2} \\
\cmidrule{2-10}
Protein  & $\mu_{10}$ & $\tau_4$  & $\tau_8$   & $\mu_{10}$ & $\tau_4$  & $\tau_8$   & $\mu_{10}$ & $\tau_4$  & $\tau_8$   \\
\midrule
1OCC   & \SI{3.3}{\angstrom} & \SI{2.0}{\percent} & \SI{42.4}{\percent} & \SI{1.9}{\angstrom} & \SI{5.6}{\percent} & \SI{52.6}{\percent} & \SI{2.0}{\angstrom} & \SI{5.4}{\percent} & \SI{51.0}{\percent} \\
1PV6   & \SI{5.3}{\angstrom} & \SI{0.0}{\percent} & \SI{8.3}{\percent}  & \SI{4.3}{\angstrom} & \SI{0.0}{\percent} & \SI{35.9}{\percent} & \SI{4.2}{\angstrom} & \SI{0.0}{\percent} & \SI{34.6}{\percent} \\
1PY6   & \SI{4.2}{\angstrom} & \SI{0.0}{\percent} & \SI{19.8}{\percent} & \SI{3.5}{\angstrom} & \SI{0.0}{\percent} & \SI{27.7}{\percent} & \SI{3.3}{\angstrom} & \SI{0.6}{\percent} & \SI{32.7}{\percent} \\
1RHZ   & \SI{4.7}{\angstrom} & \SI{0.0}{\percent} & \SI{5.5}{\percent}  & \SI{3.3}{\angstrom} & \SI{0.7}{\percent} & \SI{22.2}{\percent} & \SI{3.5}{\angstrom} & \SI{0.4}{\percent} & \SI{24.0}{\percent} \\
\bottomrule
\end{tabular}

  \caption[Sampling accuracy is improving with an increasing number of EPR
  restraints]{\textbf{Sampling accuracy is improving with an increasing number of \gls{epr}
      restraints.} The percentages of models sampled with \gls{rmsd100} values less than
    \SIlist{4;8}{\angstrom} ($\tau_4$ and $\tau_8$) are increasing with the number of restraints
    increase from one distance restraint per ten residues within \glspl{sse} to one restraints per
    three residues within \glspl{sse} to one restraint per two residues within \glspl{sse}. An
    upper limit is met at one restraint per three residues for 1OCC, 1PV6, and 1RHZ since the
    further accuracy improvements would require a more effective sampling of possible dihedral
    angle conformations.}
  \label{tab:mpepr_number_restraints}
\end{table}

\begin{figure}
  \centering
  \includegraphics[width=\textwidth]{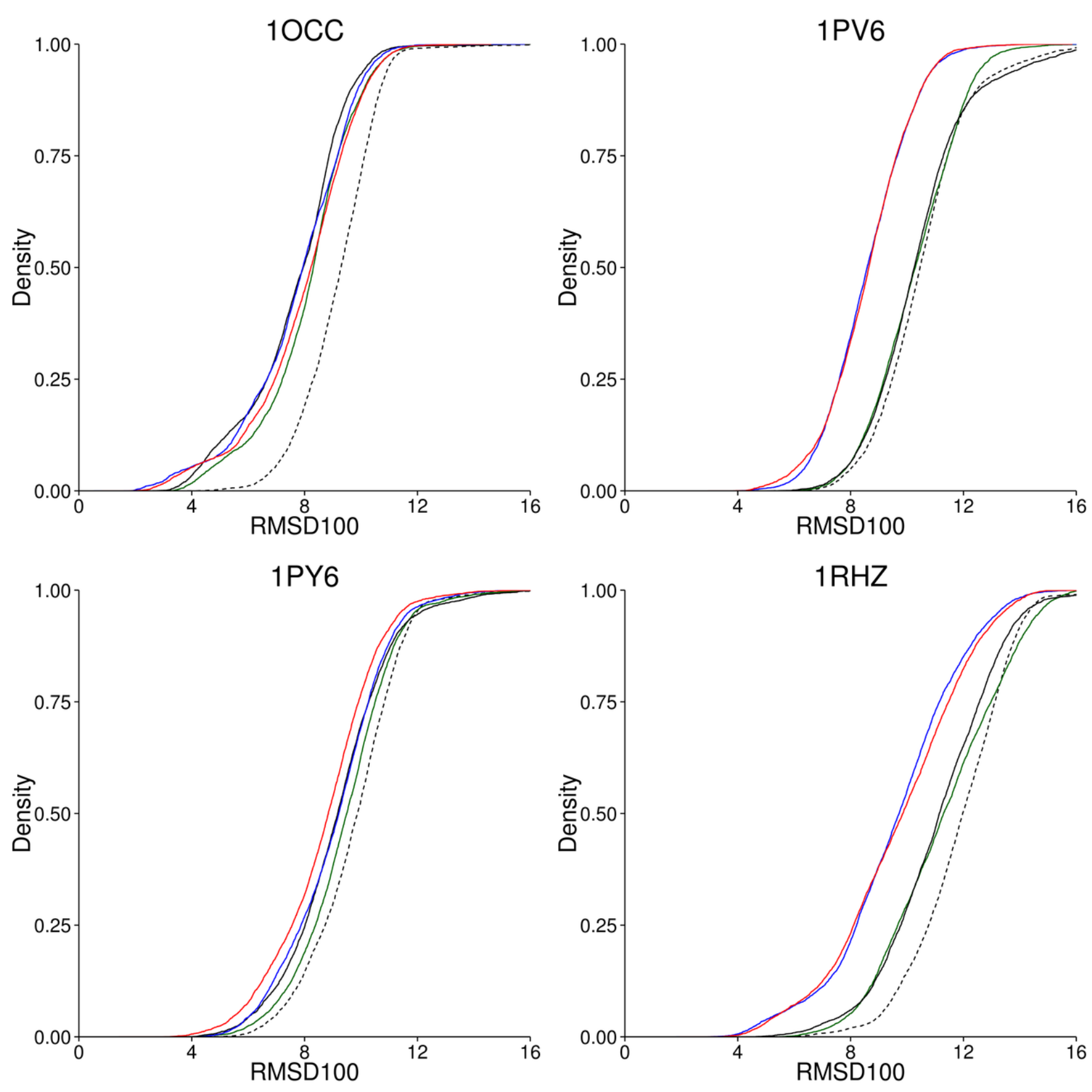}
  \caption[Influence of the number of EPR restraints on the prediction accuracy]{\textbf{Influence
      of the number of \gls{epr} restraints on the prediction accuracy.}The tertiary structure of
    four proteins was predicted with varying numbers of \gls{epr} distance restraints. Without
    restraints (dashed black), one restraint per ten residues with \glspl{sse} (green), one
    restraint per five residues (solid black), one restraint per three residues (blue), and one
    restraint per two residues (red). Shown is the cumulative density (y-axis) of models with
    respect to their \gls{rmsd100} values (x-axis).}
  \label{fig:mpepr_number_restraints}
\end{figure}

For four proteins, the influence of varying numbers of restraints was examined. In addition to the
one restraint per five residues within \glspl{sse} setup used for all benchmark cases, the tertiary
structure of 1OCC, 1PV6, 1PY6, and 1RHZ was predicted using one restraint per ten residues, one
restraint per three residues, and one restraint per two residues within \glspl{sse}. For 1PY6, the
sampling accuracy could be steadily improved with an increasing number of restraints demonstrated
by $\tau_8$ values increasing from \SI{15}{\percent} to \SI{20}{\percent} to \SI{24}{\percent} to
\SI{28}{\percent} to \SI{33}{\percent} and $\mu_{10}$ values improving from \SI{4.4}{\angstrom} to
\SI{4.2}{\angstrom} to \SI{3.6}{\angstrom} to \SI{3.5}{\angstrom} to \SI{3.3}{\angstrom} for
structure prediction without restraints, one restraint per ten residues, one restraint per five
residues, one restraint per three residues and one restraint per two residues (see
\fref{tab:mpepr_number_restraints} and \fref{fig:mpepr_number_restraints}). For 1OCC, 1PV6, and
1RHZ, a significant improvement in sampling accuracy is observed for using one restraint per three
residues instead of one restraint per ten residues within \glspl{sse}, which is demonstrated by
improvements in $\tau_8$ values from \SI{42}{\percent} to \SI{53}{\percent}, from \SI{8}{\percent}
to \SI{36}{\percent}, and from \SI{6}{\percent} to \SI{22}{\percent} and by improvements in
$\mu_{10}$ values from \SI{3.2}{\angstrom} to \SI{1.9}{\angstrom}, from \SI{5.3}{\angstrom} to
\SI{4.3}{\angstrom}, and from \SI{4.7}{\angstrom} to \SI{3.3}{\angstrom}, respectively. Increasing
the number of restraints to one restraint per two residues within \glspl{sse} fails to further
improve the sampling accuracy. We attribute this observation to significant bends in some of the
\glspl{sse} that are currently not sampled sufficiently dense by BCL::MP-Fold.

\subsection{Using experimentally obtained \gls{epr} distance restraints for rhodopsin}

The benchmark was extended to also contain rhodopsin (\Gls{pdb} entry 1GZM) for which \gls{epr}
distance measurements were available \autocite{Altenbach2008}. Although only sixteen \gls{epr}
distance restraints were available, which amounts to less than one restraint per ten residues
within \glspl{sse}, the sampling accuracy as well as the enrichment improve significantly. The
$\mu_{10}$ values improved from \SI{4.9}{\angstrom} for folding without restraints to
\SI{4.4}{\angstrom} when using restraints. The enrichment values could be improved from \num{0.6}
to \num{1.2} demonstrating that even a small number of restraints improves discrimination of
incorrect models.

\begin{figure}
  \centering
  \includegraphics[height=0.78\textheight]{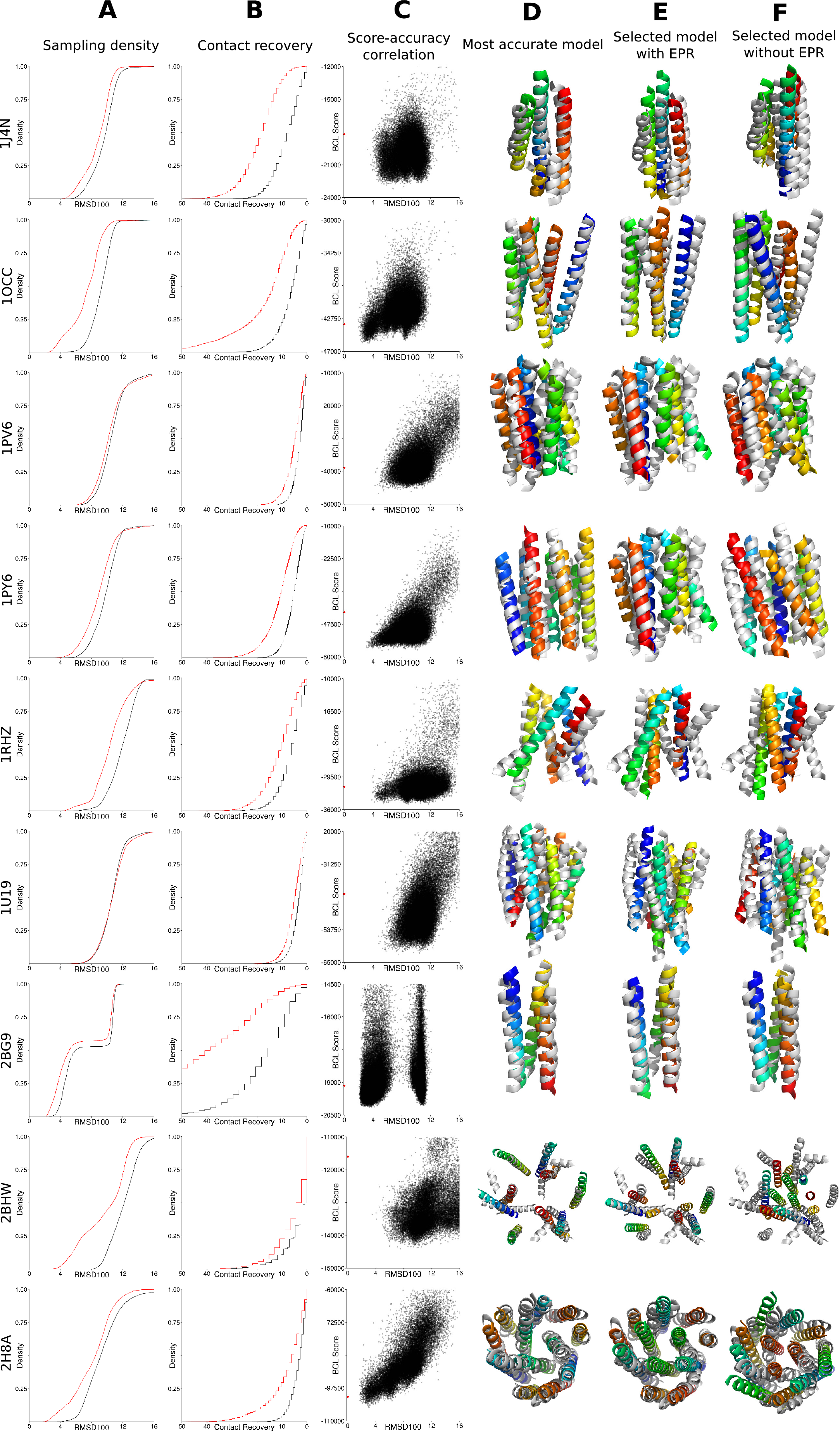}
  \caption[Gallery of the structure prediction results when using EPR data]{\textbf{Gallery of the
      structure prediction results when using \gls{epr} data.} By using \gls{epr} distance and
    accessibility restraints, the sampling accuracy is significantly improved as the selection
    ability regarding accurate models. For selected proteins, a comparison of the \gls{rmsd100}
    (column A) and Contact Recovery (column B) distributions for sampling with (red) and without
    (black) \gls{epr} restraints is shown. The y-axis of column A shows the cumulative density of
    models with respect to the \gls{rmsd100}. The y-axis of column B shows the cumulative density
    of models with respect to their contact recovery. Column C shows the correlation between the
    BCL score and the \gls{rmsd100} for the models sampled with \gls{epr} restraints (black dots)
    and the experimentally determined structure (red dot). The y-axis is the pseudo-energy score
    the algorithm assigned to the structure; the x-axis is the \gls{rmsd100} relative to the
    experimentally determined structure. The superimpositions show the best models by \gls{rmsd100}
    for folding with \gls{epr} restraints (column D), the best model by pseudo-energy score for
    folding with \gls{epr} restraints (column E), and the best model by pseudo-energy score for
    folding without \gls{epr} restraints (column F) superimposed with the experimentally determined
    structure (grey).}
  \label{fig:mpepr_gallery}
\end{figure}

\section{Discussion}

\Gls{epr} distance and accessibility restraints can aid the prediction of membrane protein
structure. For this purpose, \gls{epr} specific scores were coupled with the protein structure
prediction method BCL::MP-Fold. BCL::MP-Fold assembles predicted \glspl{sse} in space without
explicitly modeling the \gls{sse} connecting loop regions. This allows for rapid sampling of
complex topology that is not easily achieved when an intact protein backbone must be maintained. By
adding \gls{epr} specific scores to the knowledge-based scoring function, sampling of accurate
structures is increased. Additionally the selection of the most accurate models could be improved
significantly.

However, it has to be clearly stated that --- with the exception of bovine rhodopsin (\Gls{pdb}
entry 1GZM) --- all \gls{epr} restraints used in this study were simulated using the \gls{cone}
model. Therefore, the relevance of our findings depends on how well the \gls{cone} model describes
the nature of experimental \gls{deer} measurements and in particular the mobility of the spin
label.

\subsection{\Gls{epr} distance scores improve the accuracy of topologies predicted for membrane proteins}

\Gls{epr} distance measurements are associated with large uncertainties in relating the measured
spin label – spin label distance into backbone distances. In spite of this, \gls{epr} distance
measurements provide important data on membrane protein structures \autocite{Zou2009,
  Altenbach2008, Claxton2010, Zou2009}. In the present study, it has been demonstrated that
\gls{epr} distance data can significantly increase the frequency with which the correct topology of
a membrane protein is sampled (\Fref{fig:mpepr_results} and \fref{fig:mpepr_gallery}). This is
important because as the correct topologies are sampled with higher accuracy, models start to reach
the point where they can be subjected to atomic detail refinement to further increase their
accuracy \autocite{Barth2007}.

It is crucial to distinguish between the two major challenges in \emph{de novo} structure
prediction --- sampling and scoring: The average improvement in sampling accuracy ---
\emph{i.e.}~the best model built among \num{5000} independent folding trajectories --- of
\SI{0.8}{\angstrom} is moderate but significant. However, inclusion of the \gls{epr} data does not
only allow folding of models that are more accurate, it greatly improves discrimination of
incorrect models with a scoring function that combines BCL knowledge-based potentials and \gls{epr}
restraints. Without using \gls{epr} restraints the average enrichment is \num{1.3},
\emph{i.e.}~\SI{13}{\percent} of the most accurate models are in a sample of \SI{10}{\percent} best
scoring models, which is close to chance. By using \gls{epr} data in addition to the
knowledge-based score enrichment increases to \num{2.5}, \emph{i.e.}~one out of four models in the
\SI{10}{\percent} best scoring models also has the correct fold. This is important as it greatly
improves the chance to identify correctly folded models, e.g. through clustering of good-scoring
models. The combination of improved sampling and discrimination thereby significantly improves the
reliability with which were able to predict the tertiary structure of a protein.

The \gls{epr} distance data used for the present study is simulated from known experimental
structures. It will be interesting to repeat this benchmark once sufficiently dense experimental
data sets for several membrane proteins become available. For now, considerable effort was put
forth to ensure that the simulated data mimics what would be obtained from a true \gls{epr}
experiment, so that any results are unbiased by the simulated data. The previously published method
for selecting distance restraints was used to create ten different data sets per protein
\autocite{Kazmier2011}. This ensures results are not biased by a particularly selected data
set. Previously, the uncertainty in the difference between spin label distances and the
corresponding $\mathrm{C_\beta}$ distance ($D_{\mathit{SL}} - D_{\mathit{BB}}$) was accounted for
in simulated distance restraints by adding a random value between \SI{12.5}{\angstrom} and
\SI{-2.5}{\angstrom} \autocite{Kazmier2011}. Here, the probability of observing a given
$D_{\mathit{SL}} - D_{\mathit{BB}}$ is used to determine the amount that should be added to the
$\mathrm{C_\beta}$-$\mathrm{C_\beta}$ distance measured from the experimental structure.

Using a method developed for soluble proteins to select restraints for membrane proteins is not
necessarily ideal. The constraints already imposed upon membrane proteins by the membrane geometry
suggest that optimized methods for selecting restraints for membrane proteins should be
developed. One such strategy could be to measure distances between transmembrane segments on the
same side of the membrane, with the assumption that transmembrane helices are mostly rigid,
parallel structures. Further, additional work is needed to account for topologically important
\glspl{sse} that do not span the membrane, as well take into account the deviations of
transmembrane segments from ideal geometries.

The improved sampling accuracy in the protein structure prediction process is primarily caused by
the distance restraints. Whereas by using \gls{epr} accessibility restraints the average $\mu_{10}$
value over all twenty-nine proteins drops from \SI{6.0}{\angstrom} to \SI{5.8}{\angstrom}, by using
\gls{epr} distance restraints the average $\mu_{10}$ value could be improved to
\SI{5.1}{\angstrom}.

\subsection{Why not use the membrane depth parameter as additional restraint?}

Of note is that \gls{epr}-derived accessibility measurements have been also used to the determine
membrane depth parameter $\Phi$ \autocite{Altenbach1994, Frazier2002, Nielsen2005}. For this
purpose, the accessibility $\Pi$ of a single residue to two paramagnetic reagents are compared, the
water-soluble (nickel-(II)-ethylenediaminediacetate --- NiEDDA) and the
membrane-soluble (molecular oxygen --- $O_2$). The ratio of both values is used to compute the
membrane depth parameter: $\Phi_n = ln(\Pi_{O_2} / \Pi_{NiEDDA})$. The present approach
does not test effectiveness of a score that relies on the membrane depth parameter for membrane
protein structure prediction for several reasons:
\begin{inparaenum}[a)]
\item we hypothesize that knowledge-based potentials will be capable of placing transmembrane
  \glspl{sse} at the right depth for this placement should again be dominated by polarity which is
  well captured in such potentials (read above), and
\item the membrane depth parameter $\Phi_n$ is affiliated with a larger error margin for
  NiEDDA accessibilities become very small in the core of the membrane and they omit averaging
  over multiple residues.
\end{inparaenum}
Nevertheless, testing if a membrane depth related score can improve BCL::MP-Fold could be a goal in
a future experiment.

\subsection{Improved secondary structure predictions will improve the accuracy of predicted
  structures}

The \gls{sse} pools are created in order to reduce the possibility of missing a \gls{sse}, which is
generally a successful approach as demonstrated previously for soluble proteins
\autocite{Karakas2012}. The helical transmembrane span prediction software \gls{octopus}
\autocite{Viklund2008} is used in conjunction with Jufo9D \autocite{Leman2013}. Jufo9D provides
predictions for \glspl{sse} that do not necessarily span the membrane and therefore will not be
predicted by \gls{octopus}. Improved secondary structure prediction methods will benefit membrane
protein structure prediction. In addition, it has been demonstrated that the pattern of
accessibility values for measurements along a sequence follow the periodicity of the \gls{sse} on
which they are measured \autocite{Lietzow2004, Zou2009a, Zou2009}. Measured accessibility profiles
could therefore be used to inform the pool of \glspl{sse} used for structure prediction.

The pool of \glspl{sse} used to assemble the membrane protein topologies is the most important
determinant in successfully predicting the membrane proteins’ structure. This is seen for 1U19 and
2BL2. With predicted \glspl{sse}, the structure of the two proteins can be sampled to $\mu_{10}$
values of \SIlist{5.9;6.2}{\angstrom}, respectively (\Fref{tab:mpepr_sampling}). By using \gls{sse}
definitions extracted from the experimentally determined structure, the proteins can be sampled at
$\mu_{10}$ values of \SIlist{4.4;2.6}{\angstrom}, respectively. This is caused by secondary
structure prediction methods breaking up transmembrane helices into several short helices making it
harder to assemble the tertiary structure that does not have loop going through the membrane. The
experiment was repeated with \gls{sse} definitions obtained from the experimentally determined
structures of the proteins. Whereas with predicted \glspl{sse} average $\mu_{10}$, $\tau_4$, and
$\tau_8$ values of \SI{5.0}{\angstrom}, \SI{6}{\percent}, and \SI{19}{\percent} are achieved over
all twenty-nine proteins, by using the \gls{sse} definitions from the experimentally determined
structure we could improve them to \SI{4.5}{\angstrom}, \SI{8}{\percent}, and \SI{25}{\percent}. In
twenty-one out of twenty-nine cases the average accuracy of the ten best models by \gls{rmsd100}
could be improved by using \gls{sse} definitions obtained from the experimentally determined
structure (\Fref{fig:mpepr_results}). This demonstrates that further improvements of the secondary
structure prediction will also lead to an improved sampling accuracy of BCL::Fold.

\begin{wrapfigure}{R}{0.5\textwidth}
  \centering
  \includegraphics[width=0.48\textwidth]{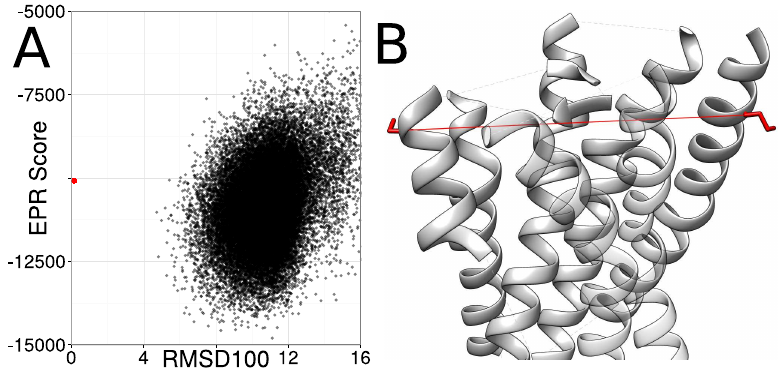}
  \caption[Limitations of the CONE model]{\textbf{Limitations of the \gls{cone} model.} For 1U19,
    the most accurate model cannot be reliably selected (A). One reason for that is, that the
    translation from the observed spin-spin distance to the backbone distance is inaccurate
    resulting in models which deviate topologically from the experimentally determined structure
    achieving a better agreement with the \gls{epr} distance restraints than the experimentally
    determined structure (B). This is demonstrated by the plot showing the correlation between the
    agreement with the \gls{epr} distance restraints (y-axis) and the \gls{rmsd100} relative to the
    experimentally determined structure (x-axis). The \gls{epr} potential does not take the
    exposure of the spin labeling site and the orientation of the
    $\mathrm{C_\alpha}$-$\mathrm{C_\beta}$ vectors into account leading to inaccuracies when
    translating $D_{\mathit{SL}}$ into $D_{\mathit{BB}}$ for the residues \numlist{7;170} of
    1U19. Both spin labels are at the outside of the protein and on different sides of the
    structure leading to greater difference between $D_{\mathit{SL}}$ and $D_{\mathit{BB}}$.}
  \label{fig:mpepr_cone_limitations}
\end{wrapfigure}

\subsection{Limitations of the \gls{cone} model knowledge-based potential}

The unknown label conformation is taken into account by the \gls{cone} model, which yields a
$D_{\mathit{SL}} - D_{\mathit{BB}}$ distribution. This wide probability distribution accounts for
two inherently different aspects --- a structural and a dynamical: The structural effect looks at
the relative position of the unpaired electron with respect to the protein backbone. This
positioning is dependent on the protein structure, specifically the direction in which the
$\mathrm{C_\alpha}$-$\mathrm{C_\beta}$ vector project into space with respect to the
$\mathrm{C_\alpha}$-$\mathrm{C_\alpha}$ vector that links the two labeling site. As the \gls{cone}
model is applied in a model-independent fashion, it does not consider these geometric features but
expresses the resulting ambiguity as part of the probability distribution. Second, chemical
environment and exposure cause variable levels of spin label dynamics. These result in distance
distributions of variable tightness in \gls{epr} experiments. This information is currently not
considered as parameter in the \gls{cone} model but absorbed by using a very wide
$D_{\mathit{SL}} - D_{\mathit{BB}}$ probability distribution. This approach has the advantage that
it is very robust with respect to uncertainties within the \gls{epr} experimental parameters and
very fast to compute. At the same time, the \gls{cone} model knowledge-based potential neglects
important geometric parameters. Developing and testing approaches that take these parameters into
account and lead to tighter distance distributions without losing the advantages of speed and
robustness is an active area of our research.

Not considering geometrical features hinders the selection of accurate models for 1U19. \Gls{epr}
distance restraints improved the sampling accuracy, but it is still not possible to reliably select
accurate models (\Fref{fig:mpepr_cone_limitations}). Although the distances observed in \gls{epr}
experiments are typically long and therefore allow a broad range of topologically different models
to fulfill them, inaccuracies in the translation from $D_{\mathit{SL}}$ to $D_{\mathit{BB}}$ also
contribute to the selection problem. In the case of 1U19 the experimentally determined structure,
which served as the template for the simulation of the \gls{epr} distance restraints, shows a worse
agreement with the restraints than the best scoring models. The spin-spin distance between residue
\num{7} and residue \num{170} is \SI{43.6}{\angstrom}, whereas the distance between the
$\mathrm{C_\beta}$-atoms is \SI{35.7}{\angstrom} resulting in an agreement score of \num{0.3} on a
scale from \numrange{0}{1}. Following the \gls{epr} potential, a
$\mathrm{C_\beta}$-$\mathrm{C_\beta}$ distance of \SI{41.1}{\angstrom} is favorable, which is
accomplished by the sampled models with the best score leading to the selection of models, which
deviate significantly from the experimentally determined structure. Both spin labeling sites are
exposed, indicating they are at the outside of the protein. The projection angle between the
$\mathrm{C_\alpha}$-$\mathrm{C_\beta}$ vectors is greater than \SI{160}{\degree}, making it more
likely that the spin labels are pointing away from each other. Those two properties allow the
inference that we would expect a larger difference between $D_{\mathit{SL}}$ and $D_{\mathit{BB}}$
than \SI{2.5}{\angstrom}. By using a knowledge-based potential, which also takes the exposure of
the spin labeling sites and additional geometrical information into account a better ranking of the
sampled models would be possible.

\subsection{Ambiguities in the ranking of models remain}

Although the usage of restraints obtained from \gls{epr} experiments significantly improves the
discrimination of incorrect models, ambiguities in the ranking of the models remain for multiple
proteins in the benchmark set. This observation was especially pronounced for the proteins 1J4N,
1PV6, 1PY6, and 1U19 (\Fref{fig:mpepr_gallery}). In those cases, the best \SI{10}{\percent} of the
models by BCL score cover a wide range of topologies. For 1PV6, the best \SI{10}{\percent} of the
models by BCL score cover an \gls{rmsd100} range of \SI{8}{\angstrom} when compared to the
experimentally determined structure. Multiple factors are contributing to this observation. First,
the BCL::Fold scoring function is an inaccurate approximation of free energy, which limits its
discriminative power \autocite{Wotzel2012}. Although adding a term that measures agreement with
experimental data will improve its discriminative power, it appears that sparse restraints from
\gls{epr} data are sometimes insufficient to remove all ambiguities. This is also because, second,
the translation of spin label distance distributions into a backbone structural restraint
introduces a substantial uncertainty and therefore allows sometimes multiple topologies to fulfill
the restraint. One side effect of these approximations is that --- as shown in
\fref{fig:mpepr_gallery} --- the native structure is not always in the global minimum of the BCL
scoring function. Relaxing the experimentally determined protein structures in the BCL force field
indicate that the closest minimum in the scoring function is between \SI{1.5}{\angstrom} and
\SI{4.1}{\angstrom} in \gls{rmsd100} separate relative to the experimentally determined structures.

\section{Conclusion}

The determination of membrane protein folds from \gls{epr} distance and accessibility data is
within reach if these restraints aid protein folding protocols such as BCL::MP-Fold. The ability of
\gls{epr} data to improve the sampling of native-like topologies and the importance of \gls{epr}
accessibility data for obtaining highest contact recovery values was demonstrated. Further, the
\gls{epr} specific scores allow the selection of close-to-native models, thereby overcoming a major
obstacle in \emph{de novo} protein structure prediction. Refining \gls{epr} distance potentials to
also take the exposure of the spin labeling sites as well as relative orientation of the
$\mathrm{C_\alpha}$-$\mathrm{C_\beta}$ vector might provide a more accurate translation from
spin-spin distance into backbone distance, thereby further increasing model quality.

\section{Acknowledgments}

We thank Cristian Altenbach and Wayne Hubbell for sharing their \gls{epr} data for rhodopsin
(\Gls{pdb} entry 1GZM) with us and therefore enabling us to evaluate our algorithm based on
experimentally determined data.

Parts of the data analysis were performed using the R package with ggplot2. The renderings of the
models were created using Chimera. The composite figures were created using Inkscape.

This research used resources of the Oak Ridge Leadership Computing Facility
at the Oak Ridge National Laboratory, which is supported by the Office of Science of the
U.S. Department of Energy under Contract No. DE-AC05-00OR22725.


\section*{Availability}

The BCL software suite is available at \url{http://www.meilerlab.org/bclcommons} under academic and
business site licenses. The BCL source code is published under the BCL license and is available at
\url{http://www.meilerlab.org/bclcommons}.

\printbibliography

\section*{Supplementary Material}

\begin{sidewaystable}
  \small
  \centering
  \begin{tabular}{lrrrrrrrrrrrrrrrr}
\toprule
  & \multicolumn{4}{c}{None} & \multicolumn{4}{c}{Accessibility} & \multicolumn{4}{c}{Distance} & \multicolumn{4}{c}{Accessibility \& Distance} \\
\cmidrule{2-17}
Protein & $\mathit{best}$ & $\mu_{10}$  & $\tau_4$   & $\tau_8$   & $\mathit{best}$ & $\mu_{10}$ & $\tau_4$   & $\tau_8$   & $\mathit{best}$ & $\mu_{10}$ & $\tau_4$   & $\tau_8$   & $\mathit{best}$ & $\mu_{10}$ & $\tau_4$   & $\tau_8$   \\
\midrule
1IWG   & \SI{4.2}{\angstrom}  & \SI{4.9}{\angstrom}  & \SI{0.0}{\percent}  & \SI{19.1}{\percent}& \SI{3.9}{\angstrom}  & \SI{4.3}{\angstrom} & \SI{0.1}{\percent}  & \SI{26.3}{\percent} & \SI{3.8}{\angstrom}  & \SI{4.4}{\angstrom} &\SI{0.0}{\percent}  & \SI{22.1}{\percent} & \SI{3.6}{\angstrom}  & \SI{4.2}{\angstrom} & \SI{0.1}{\percent}  & \SI{25.8}{\percent}\\
1J4N   & \SI{5.2}{\angstrom}  & \SI{5.3}{\angstrom}  & \SI{0.0}{\percent}  & \SI{18.0}{\percent}& \SI{4.2}{\angstrom}  & \SI{4.7}{\angstrom} & \SI{0.0}{\percent}  & \SI{22.5}{\percent} & \SI{4.4}{\angstrom}  & \SI{4.8}{\angstrom} &\SI{0.0}{\percent}  & \SI{21.0}{\percent} & \SI{3.8}{\angstrom}  & \SI{4.3}{\angstrom} & \SI{0.0}{\percent}  & \SI{25.4}{\percent}\\
1KPL   & \SI{10.1}{\angstrom} & \SI{10.3}{\angstrom} & \SI{0.0}{\percent}  & \SI{0.0}{\percent} & \SI{8.8}{\angstrom}  & \SI{9.6}{\angstrom} & \SI{0.0}{\percent}  & \SI{0.0}{\percent}  & \SI{8.5}{\angstrom}  & \SI{9.4}{\angstrom} &\SI{0.0}{\percent}  & \SI{0.0}{\percent}  & \SI{8.5}{\angstrom}  & \SI{9.3}{\angstrom} & \SI{0.0}{\percent}  & \SI{0.0}{\percent} \\
1OCC   & \SI{3.7}{\angstrom}  & \SI{4.5}{\angstrom}  & \SI{0.0}{\percent}  & \SI{20.1}{\percent}& \SI{2.2}{\angstrom}  & \SI{2.9}{\angstrom} & \SI{0.8}{\percent}  & \SI{30.2}{\percent} & \SI{2.4}{\angstrom}  & \SI{2.8}{\angstrom} &\SI{3.1}{\percent}  & \SI{44.9}{\percent} & \SI{2.0}{\angstrom}  & \SI{2.3}{\angstrom} & \SI{7.0}{\percent}  & \SI{50.8}{\percent}\\
1OKC   & \SI{6.4}{\angstrom}  & \SI{6.8}{\angstrom}  & \SI{0.0}{\percent}  & \SI{1.2}{\percent} & \SI{8.2}{\angstrom}  & \SI{8.8}{\angstrom} & \SI{0.0}{\percent}  & \SI{0.0}{\percent}  & \SI{7.1}{\angstrom}  & \SI{7.8}{\angstrom} &\SI{0.0}{\percent}  & \SI{0.1}{\percent}  & \SI{6.9}{\angstrom}  & \SI{7.7}{\angstrom} & \SI{0.0}{\percent}  & \SI{0.2}{\percent} \\
1PV6   & \SI{5.2}{\angstrom}  & \SI{6.1}{\angstrom}  & \SI{0.0}{\percent}  & \SI{5.3}{\percent} & \SI{5.3}{\angstrom}  & \SI{5.8}{\angstrom} & \SI{0.0}{\percent}  & \SI{6.0}{\percent}  & \SI{5.2}{\angstrom}  & \SI{5.7}{\angstrom} &\SI{0.0}{\percent}  & \SI{8.5}{\percent}  & \SI{4.9}{\angstrom}  & \SI{5.5}{\angstrom} & \SI{0.0}{\percent}  & \SI{10.5}{\percent}\\
1PY6   & \SI{4.4}{\angstrom}  & \SI{5.1}{\angstrom}  & \SI{0.0}{\percent}  & \SI{15.2}{\percent}& \SI{3.9}{\angstrom}  & \SI{4.3}{\angstrom} & \SI{0.0}{\percent}  & \SI{17.5}{\percent} & \SI{3.6}{\angstrom}  & \SI{4.2}{\angstrom} &\SI{0.1}{\percent}  & \SI{24.2}{\percent} & \SI{3.3}{\angstrom}  & \SI{3.8}{\angstrom} & \SI{0.2}{\percent}  & \SI{26.7}{\percent}\\
1RHZ   & \SI{5.6}{\angstrom}  & \SI{5.9}{\angstrom}  & \SI{0.0}{\percent}  & \SI{2.1}{\percent} & \SI{5.5}{\angstrom}  & \SI{5.7}{\angstrom} & \SI{0.0}{\percent}  & \SI{1.9}{\percent}  & \SI{4.2}{\angstrom}  & \SI{4.7}{\angstrom} &\SI{0.0}{\percent}  & \SI{5.6}{\percent}  & \SI{3.9}{\angstrom}  & \SI{4.4}{\angstrom} & \SI{0.0}{\percent}  & \SI{7.0}{\percent} \\
1U19   & \SI{5.3}{\angstrom}  & \SI{5.7}{\angstrom}  & \SI{0.0}{\percent}  & \SI{4.7}{\percent} & \SI{5.4}{\angstrom}  & \SI{6.1}{\angstrom} & \SI{0.0}{\percent}  & \SI{3.0}{\percent}  & \SI{5.5}{\angstrom}  & \SI{6.2}{\angstrom} &\SI{0.0}{\percent}  & \SI{3.6}{\percent}  & \SI{5.2}{\angstrom}  & \SI{5.9}{\angstrom} & \SI{0.0}{\percent}  & \SI{4.1}{\percent} \\
1XME   & \SI{8.0}{\angstrom}  & \SI{8.7}{\angstrom}  & \SI{0.0}{\percent}  & \SI{0.0}{\percent} & \SI{8.1}{\angstrom}  & \SI{8.7}{\angstrom} & \SI{0.0}{\percent}  & \SI{0.0}{\percent}  & \SI{7.3}{\angstrom}  & \SI{7.8}{\angstrom} &\SI{0.0}{\percent}  & \SI{0.2}{\percent}  & \SI{7.2}{\angstrom}  & \SI{7.8}{\angstrom} & \SI{0.0}{\percent}  & \SI{0.2}{\percent} \\
2BG9   & \SI{2.3}{\angstrom}  & \SI{2.6}{\angstrom}  & \SI{10.8}{\percent} & \SI{52.7}{\percent}& \SI{2.2}{\angstrom}  & \SI{2.3}{\angstrom} & \SI{28.1}{\percent} & \SI{51.3}{\percent} & \SI{2.2}{\angstrom}  & \SI{2.3}{\angstrom} &\SI{28.1}{\percent} & \SI{57.7}{\percent} & \SI{2.1}{\angstrom}  & \SI{2.2}{\angstrom} & \SI{32.6}{\percent} & \SI{61.3}{\percent}\\
2BL2   & \SI{7.5}{\angstrom}  & \SI{8.4}{\angstrom}  & \SI{0.0}{\percent}  & \SI{0.0}{\percent} & \SI{7.8}{\angstrom}  & \SI{8.4}{\angstrom} & \SI{0.0}{\percent}  & \SI{0.0}{\percent}  & \SI{5.0}{\angstrom}  & \SI{6.1}{\angstrom} &\SI{0.0}{\percent}  & \SI{0.9}{\percent}  & \SI{5.1}{\angstrom}  & \SI{6.2}{\angstrom} & \SI{0.0}{\percent}  & \SI{0.8}{\percent} \\
2BS2   & \SI{6.1}{\angstrom}  & \SI{6.3}{\angstrom}  & \SI{0.0}{\percent}  & \SI{2.4}{\percent} & \SI{6.3}{\angstrom}  & \SI{6.6}{\angstrom} & \SI{0.0}{\percent}  & \SI{1.8}{\percent}  & \SI{5.3}{\angstrom}  & \SI{5.9}{\angstrom} &\SI{0.0}{\percent}  & \SI{3.5}{\percent}  & \SI{5.2}{\angstrom}  & \SI{5.8}{\angstrom} & \SI{0.0}{\percent}  & \SI{3.5}{\percent} \\
2IC8   & \SI{5.5}{\angstrom}  & \SI{6.1}{\angstrom}  & \SI{0.0}{\percent}  & \SI{4.7}{\percent} & \SI{6.0}{\angstrom}  & \SI{6.2}{\angstrom} & \SI{0.0}{\percent}  & \SI{4.2}{\percent}  & \SI{5.0}{\angstrom}  & \SI{5.5}{\angstrom} &\SI{0.0}{\percent}  & \SI{9.8}{\percent}  & \SI{5.1}{\angstrom}  & \SI{5.5}{\angstrom} & \SI{0.0}{\percent}  & \SI{9.4}{\percent} \\
2K73   & \SI{3.3}{\angstrom}  & \SI{3.6}{\angstrom}  & \SI{0.5}{\percent}  & \SI{24.8}{\percent}& \SI{3.3}{\angstrom}  & \SI{3.5}{\angstrom} & \SI{0.7}{\percent}  & \SI{21.4}{\percent} & \SI{2.7}{\angstrom}  & \SI{3.0}{\angstrom} &\SI{3.4}{\percent}  & \SI{30.1}{\percent} & \SI{2.8}{\angstrom}  & \SI{3.1}{\angstrom} & \SI{3.3}{\percent}  & \SI{30.0}{\percent}\\
2KSF   & \SI{4.1}{\angstrom}  & \SI{4.4}{\angstrom}  & \SI{0.0}{\percent}  & \SI{22.3}{\percent}& \SI{2.6}{\angstrom}  & \SI{3.0}{\angstrom} & \SI{2.6}{\percent}  & \SI{21.9}{\percent} & \SI{2.9}{\angstrom}  & \SI{3.1}{\angstrom} &\SI{5.2}{\percent}  & \SI{25.9}{\percent} & \SI{2.7}{\angstrom}  & \SI{2.9}{\angstrom} & \SI{8.3}{\percent}  & \SI{25.1}{\percent}\\
2KSY   & \SI{4.9}{\angstrom}  & \SI{5.3}{\angstrom}  & \SI{0.0}{\percent}  & \SI{11.2}{\percent}& \SI{3.7}{\angstrom}  & \SI{4.5}{\angstrom} & \SI{0.0}{\percent}  & \SI{13.4}{\percent} & \SI{3.5}{\angstrom}  & \SI{4.1}{\angstrom} &\SI{0.1}{\percent}  & \SI{21.4}{\percent} & \SI{3.6}{\angstrom}  & \SI{4.1}{\angstrom} & \SI{0.1}{\percent}  & \SI{20.4}{\percent}\\
2NR9   & \SI{5.6}{\angstrom}  & \SI{6.0}{\angstrom}  & \SI{0.0}{\percent}  & \SI{5.8}{\percent} & \SI{6.0}{\angstrom}  & \SI{6.6}{\angstrom} & \SI{0.0}{\percent}  & \SI{3.4}{\percent}  & \SI{6.4}{\angstrom}  & \SI{6.9}{\angstrom} &\SI{0.0}{\percent}  & \SI{2.5}{\percent}  & \SI{6.4}{\angstrom}  & \SI{6.9}{\angstrom} & \SI{0.0}{\percent}  & \SI{2.5}{\percent} \\
2XUT   & \SI{7.7}{\angstrom}  & \SI{8.5}{\angstrom}  & \SI{0.0}{\percent}  & \SI{0.0}{\percent} & \SI{7.5}{\angstrom}  & \SI{8.4}{\angstrom} & \SI{0.0}{\percent}  & \SI{0.0}{\percent}  & \SI{7.7}{\angstrom}  & \SI{8.2}{\angstrom} &\SI{0.0}{\percent}  & \SI{0.1}{\percent}  & \SI{7.7}{\angstrom}  & \SI{8.2}{\angstrom} & \SI{0.0}{\percent}  & \SI{0.1}{\percent} \\
3GIA   & \SI{10.0}{\angstrom} & \SI{10.2}{\angstrom} & \SI{0.0}{\percent}  & \SI{0.0}{\percent} & \SI{9.5}{\angstrom}  & \SI{9.8}{\angstrom} & \SI{0.0}{\percent}  & \SI{0.0}{\percent}  & \SI{9.1}{\angstrom}  & \SI{9.6}{\angstrom} &\SI{0.0}{\percent}  & \SI{0.0}{\percent}  & \SI{9.1}{\angstrom}  & \SI{9.6}{\angstrom} & \SI{0.0}{\percent}  & \SI{0.0}{\percent} \\
3KCU   & \SI{7.0}{\angstrom}  & \SI{7.5}{\angstrom}  & \SI{0.0}{\percent}  & \SI{0.3}{\percent} & \SI{7.7}{\angstrom}  & \SI{8.1}{\angstrom} & \SI{0.0}{\percent}  & \SI{0.0}{\percent}  & \SI{6.3}{\angstrom}  & \SI{7.2}{\angstrom} &\SI{0.0}{\percent}  & \SI{0.6}{\percent}  & \SI{6.5}{\angstrom}  & \SI{7.3}{\angstrom} & \SI{0.0}{\percent}  & \SI{0.4}{\percent} \\
3KJ6   & \SI{6.6}{\angstrom}  & \SI{7.0}{\angstrom}  & \SI{0.0}{\percent}  & \SI{0.7}{\percent} & \SI{4.9}{\angstrom}  & \SI{6.4}{\angstrom} & \SI{0.0}{\percent}  & \SI{1.0}{\percent}  & \SI{5.2}{\angstrom}  & \SI{5.9}{\angstrom} &\SI{0.0}{\percent}  & \SI{3.0}{\percent}  & \SI{4.9}{\angstrom}  & \SI{5.8}{\angstrom} & \SI{0.0}{\percent}  & \SI{3.0}{\percent} \\
3P5N   & \SI{4.6}{\angstrom}  & \SI{5.8}{\angstrom}  & \SI{0.0}{\percent}  & \SI{5.5}{\percent} & \SI{5.3}{\angstrom}  & \SI{5.8}{\angstrom} & \SI{0.0}{\percent}  & \SI{4.5}{\percent}  & \SI{4.8}{\angstrom}  & \SI{5.6}{\angstrom} &\SI{0.0}{\percent}  & \SI{10.2}{\percent} & \SI{4.8}{\angstrom}  & \SI{5.6}{\angstrom} & \SI{0.0}{\percent}  & \SI{10.0}{\percent}\\
\midrule
2BHW   & \SI{7.4}{\angstrom}  & \SI{7.8}{\angstrom}  & \SI{0.0}{\percent}  & \SI{0.2}{\percent} & \SI{6.9}{\angstrom}  & \SI{7.4}{\angstrom} & \SI{0.0}{\percent}  & \SI{0.3}{\percent}  & \SI{3.0}{\angstrom}  & \SI{3.4}{\angstrom} &\SI{0.9}{\percent}  & \SI{31.7}{\percent} & \SI{2.9}{\angstrom}  & \SI{3.4}{\angstrom} & \SI{0.9}{\percent}  & \SI{31.2}{\percent}\\
2H8A   & \SI{3.3}{\angstrom}  & \SI{3.9}{\angstrom}  & \SI{0.1}{\percent}  & \SI{32.5}{\percent}& \SI{3.1}{\angstrom}  & \SI{3.8}{\angstrom} & \SI{0.2}{\percent}  & \SI{34.5}{\percent} & \SI{1.9}{\angstrom}  & \SI{2.1}{\angstrom} &\SI{6.4}{\percent}  & \SI{41.7}{\percent} & \SI{1.8}{\angstrom}  & \SI{2.0}{\angstrom} & \SI{7.6}{\percent}  & \SI{44.3}{\percent}\\
2HAC   & \SI{1.3}{\angstrom}  & \SI{1.4}{\angstrom}  & \SI{75.3}{\percent} & \SI{89.6}{\percent}& \SI{1.3}{\angstrom}  & \SI{1.4}{\angstrom} & \SI{83.7}{\percent} & \SI{94.8}{\percent} & \SI{1.3}{\angstrom}  & \SI{1.5}{\angstrom} &\SI{71.8}{\percent} & \SI{92.1}{\percent} & \SI{1.3}{\angstrom}  & \SI{1.5}{\angstrom} & \SI{87.0}{\percent} & \SI{96.8}{\percent}\\
2L35   & \SI{2.6}{\angstrom}  & \SI{2.9}{\angstrom}  & \SI{4.5}{\percent}  & \SI{19.8}{\percent}& \SI{1.7}{\angstrom}  & \SI{1.8}{\angstrom} & \SI{17.1}{\percent} & \SI{22.4}{\percent} & \SI{1.9}{\angstrom}  & \SI{2.1}{\angstrom} &\SI{31.0}{\percent} & \SI{48.3}{\percent} & \SI{1.9}{\angstrom}  & \SI{2.1}{\angstrom} & \SI{31.0}{\percent} & \SI{48.3}{\percent}\\
2ZY9   & \SI{4.7}{\angstrom}  & \SI{5.7}{\angstrom}  & \SI{0.0}{\percent}  & \SI{3.4}{\percent} & \SI{4.5}{\angstrom}  & \SI{5.5}{\angstrom} & \SI{0.0}{\percent}  & \SI{4.4}{\percent}  & \SI{2.7}{\angstrom}  & \SI{3.3}{\angstrom} &\SI{0.4}{\percent}  & \SI{21.4}{\percent} & \SI{2.8}{\angstrom}  & \SI{3.4}{\angstrom} & \SI{0.3}{\percent}  & \SI{20.9}{\percent}\\
3CAP   & \SI{7.4}{\angstrom}  & \SI{8.0}{\angstrom}  & \SI{0.0}{\percent}  & \SI{0.1}{\percent} & \SI{7.3}{\angstrom}  & \SI{7.9}{\angstrom} & \SI{0.0}{\percent}  & \SI{0.1}{\percent}  & \SI{4.9}{\angstrom}  & \SI{5.4}{\angstrom} &\SI{0.0}{\percent}  & \SI{5.8}{\percent}  & \SI{4.7}{\angstrom}  & \SI{5.3}{\angstrom} & \SI{0.0}{\percent}  & \SI{4.1}{\percent} \\
\midrule
Ø      & \SI{5.5}{\angstrom}  & \SI{6.0}{\angstrom}  & \SI{3.0}{\percent}  & \SI{12.5}{\percent}& \SI{5.3}{\angstrom}  & \SI{5.8}{\angstrom} & \SI{4.6}{\percent}  & \SI{13.3}{\percent} & \SI{4.6}{\angstrom}  & \SI{5.1}{\angstrom} &\SI{5.2}{\percent}  & \SI{18.0}{\percent} & \SI{4.5}{\angstrom}  & \SI{5.0}{\angstrom} & \SI{6.2}{\percent}  & \SI{19.4}{\percent}\\
\bottomrule
\end{tabular}

  \caption[Sampling accuracy comparison for folding with and without EPR
  restraints]{\textbf{Sampling accuracy comparison for folding with and without \gls{epr}
      restraints.} Results for folding the proteins based on predicted \glspl{sse} without
    restraints, with accessibility restraints only, with distance restraints only, and with
    accessibility and distance restraints. Shown are the \gls{rmsd100} of the most accurate model
    sampled (best), the average \gls{rmsd100} of the ten most accurate models ($\mu_{10}$) as well
    as the percentage of the sampled models with \gls{rmsd100} values of less than
    \SIlist{4;8}{\angstrom} ($\tau_4$ and $\tau_8$). Using protein specific data derived from
    \gls{epr} experiments significantly improves the sampling accuracy. The improvement is mainly
    caused by using \gls{epr} distance restraints; accessibility restraints have a minor effect on
    the sampling accuracy. The proteins above the separating line are monomeric proteins; below the
    separating line are multimeric proteins.}
  \label{tab:mpepr_sampling}
\end{sidewaystable}

\begin{sidewaystable}
  \small
  \centering
  \begin{tabular}{lrrrrrrrrrrrrrrrr}
\toprule
  &  \multicolumn{4}{c}{None} & \multicolumn{4}{c}{Accessibility} & \multicolumn{4}{c}{Distance} & \multicolumn{4}{c}{Accessibility \& Distance} \\
\cmidrule{2-17}
Protein & $\mathit{best}$ & $\phi_{10}$  & $\gamma_{20}$  & $\gamma_{40}$  & $\mathit{best}$ & $\phi_{10}$  & $\gamma_{20}$  & $\gamma_{40}$  & $\mathit{best}$ & $\phi_{10}$  & $\gamma_{20}$  & $\gamma_{40}$  & $\mathit{best}$ & $\phi_{10}$  & $\gamma_{20}$  & $\gamma_{40}$  \\
\midrule
1IWG   & 23.7 & 21.3 & 0.2  & 0.0  & 33.5 & 29.4 & 2.0  & 0.0  & 31.3 & 26.9 & 1.2  & 0.0  & 33.8 & 28.3 & 2.9  & 0.0  \\
1J4N   & 31.6 & 28.7 & 2.7  & 0.0  & 45.9 & 39.7 & 13.5 & 0.1  & 40.5 & 36.8 & 11.2 & 0.0  & 44.8 & 39.8 & 22.5 & 0.3  \\
1KPL   & 22.8 & 14.4 & 0.0  & 0.0  & 18.7 & 15.5 & 0.0  & 0.0  & 19.0 & 15.6 & 0.0  & 0.0  & 18.4 & 16.1 & 0.0  & 0.0  \\
1OCC   & 44.0 & 34.3 & 1.7  & 0.0  & 77.1 & 57.8 & 20.7 & 1.1  & 63.0 & 55.9 & 15.4 & 1.5  & 66.1 & 59.4 & 25.6 & 4.6  \\
1OKC   & 15.4 & 11.0 & 0.0  & 0.0  & 22.0 & 16.4 & 0.0  & 0.0  & 19.5 & 16.4 & 0.0  & 0.0  & 19.8 & 17.6 & 0.0  & 0.0  \\
1PV6   & 19.6 & 13.9 & 0.0  & 0.0  & 23.6 & 20.6 & 0.1  & 0.0  & 22.1 & 19.0 & 0.1  & 0.0  & 22.7 & 19.5 & 0.1  & 0.0  \\
1PY6   & 21.0 & 19.1 & 0.0  & 0.0  & 35.1 & 30.4 & 2.2  & 0.0  & 32.8 & 27.4 & 1.4  & 0.0  & 36.0 & 31.1 & 3.5  & 0.0  \\
1RHZ   & 33.0 & 29.2 & 1.6  & 0.0  & 38.1 & 35.1 & 6.2  & 0.0  & 41.9 & 36.0 & 7.0  & 0.1  & 41.8 & 37.2 & 11.3 & 0.2  \\
1U19   & 25.4 & 17.3 & 0.0  & 0.0  & 25.8 & 22.7 & 0.2  & 0.0  & 21.9 & 18.4 & 0.0  & 0.0  & 23.8 & 19.0 & 0.1  & 0.0  \\
1XME   & 10.0 & 8.5  & 0.0  & 0.0  & 10.6 & 9.2  & 0.0  & 0.0  & 9.6  & 8.3  & 0.0  & 0.0  & 10.1 & 8.8  & 0.0  & 0.0  \\
2BG9   & 79.5 & 73.9 & 37.5 & 5.4  & 95.5 & 90.9 & 93.0 & 52.2 & 84.1 & 78.2 & 51.2 & 16.5 & 92.5 & 87.8 & 78.3 & 40.6 \\
2BL2   & 18.3 & 14.8 & 0.0  & 0.0  & 25.7 & 20.9 & 0.1  & 0.0  & 31.7 & 23.9 & 0.2  & 0.0  & 28.7 & 23.0 & 0.3  & 0.0  \\
2BS2   & 23.9 & 21.7 & 0.3  & 0.0  & 32.5 & 28.9 & 1.0  & 0.0  & 27.6 & 24.8 & 0.4  & 0.0  & 34.2 & 28.8 & 1.2  & 0.0  \\
2IC8   & 27.4 & 21.8 & 0.2  & 0.0  & 29.9 & 26.2 & 1.0  & 0.0  & 26.2 & 23.4 & 0.3  & 0.0  & 34.5 & 29.0 & 2.1  & 0.0  \\
2K73   & 44.2 & 39.8 & 6.5  & 0.1  & 65.1 & 60.3 & 37.2 & 3.7  & 52.3 & 47.8 & 14.0 & 0.5  & 65.8 & 60.7 & 38.5 & 6.7  \\
2KSF   & 47.2 & 37.0 & 4.1  & 0.0  & 73.6 & 66.6 & 28.6 & 3.4  & 58.5 & 53.0 & 16.6 & 0.8  & 71.5 & 65.2 & 35.9 & 6.6  \\
2KSY   & 29.6 & 22.2 & 0.2  & 0.0  & 37.6 & 31.5 & 3.3  & 0.0  & 38.8 & 28.9 & 1.4  & 0.0  & 37.7 & 31.5 & 2.9  & 0.0  \\
2NR9   & 24.1 & 20.0 & 0.1  & 0.0  & 27.4 & 22.7 & 0.4  & 0.0  & 25.1 & 20.9 & 0.1  & 0.0  & 24.4 & 23.4 & 0.3  & 0.0  \\
2XUT   & 9.8  & 9.0  & 0.0  & 0.0  & 12.3 & 10.0 & 0.0  & 0.0  & 10.0 & 9.7  & 0.0  & 0.0  & 10.1 & 8.9  & 0.0  & 0.0  \\
3GIA   & 7.6  & 6.5  & 0.0  & 0.0  & 8.3  & 7.0  & 0.0  & 0.0  & 7.6  & 6.8  & 0.0  & 0.0  & 7.8  & 6.5  & 0.0  & 0.0  \\
3KCU   & 19.1 & 15.5 & 0.0  & 0.0  & 21.5 & 18.4 & 0.0  & 0.0  & 21.2 & 17.6 & 0.0  & 0.0  & 20.3 & 19.0 & 0.0  & 0.0  \\
3KJ6   & 18.4 & 15.9 & 0.0  & 0.0  & 29.5 & 22.1 & 0.1  & 0.0  & 21.7 & 19.6 & 0.1  & 0.0  & 25.4 & 21.8 & 0.2  & 0.0  \\
3P5N   & 29.8 & 21.2 & 0.1  & 0.0  & 33.0 & 29.2 & 1.2  & 0.0  & 30.9 & 25.3 & 0.2  & 0.0  & 30.9 & 30.0 & 0.9  & 0.0  \\
\midrule
2BHW   & 42.2 & 36.8 & 2.3  & 0.0  & 46.8 & 40.0 & 3.8  & 0.2  & 49.5 & 43.1 & 4.2  & 0.2  & 48.9 & 42.3 & 4.4  & 0.2  \\
2H8A   & 28.8 & 19.5 & 0.0  & 0.0  & 30.9 & 28.2 & 1.1  & 0.0  & 44.8 & 40.9 & 7.2  & 0.2  & 49.8 & 45.7 & 10.7 & 0.7  \\
2HAC   & 100.0& 98.9 & 87.0 & 80.4 & 100.0& 98.0 & 90.9 & 83.1 & 99.4 & 98.0 & 81.7 & 73.1 & 99.4 & 97.7 & 82.6 & 75.2 \\
2L35   & 53.2 & 49.2 & 11.1 & 0.5  & 76.6 & 73.8 & 53.2 & 15.2 & 49.4 & 47.3 & 26.6 & 0.9  & 76.0 & 70.6 & 79.0 & 18.9 \\
2ZY9   & 16.9 & 14.2 & 0.0  & 0.0  & 27.1 & 23.7 & 0.3  & 0.0  & 29.6 & 24.7 & 0.3  & 0.0  & 36.8 & 30.9 & 1.4  & 0.0  \\
3CAP   & 12.7 & 11.9 & 0.0  & 0.0  & 17.9 & 14.0 & 0.0  & 0.0  & 14.4 & 12.3 & 0.0  & 0.0  & 17.5 & 14.2 & 2.8  & 0.0  \\
\midrule
Ø      & 30.3 & 25.8 & 5.4  & 3.0  & 38.7 & 34.1 & 12.4 & 5.5  & 35.3 & 31.3 & 8.3  & 3.2  & 38.9 & 35   & 14.1 & 5.3  \\
\bottomrule
\end{tabular}

  \caption[Contact recovery comparison for folding with and without EPR restraints]{\textbf{Contact
      recovery comparison for folding with and without \gls{epr} restraints.} The usage of
    \gls{epr} accessibility restraints significantly increases the percentage of recovered contacts
    (amino acids that are separated by at least six residues and have a maximum Euclidean distance
    of \SI{8}{\angstrom} in the experimentally determined structure). Shown are the highest contact
    recovery achieved (best), the average contact recovery of the ten models with the highest
    contact recovery ($\phi_{10}$) and the percentage of models for which more than
    \SIlist{20;40}{\percent} of the contacts were recovered ($\gamma_{20}$ and $\gamma_{40}$). The
    proteins above the separating line are monomeric proteins; below the separating line are
    multimeric proteins.}
  \label{tab:mpepr_contact_recovery}
\end{sidewaystable}

\begin{sidewaystable}
  \small
  \centering
  \begin{tabular}{lrrrrrrrrrrrrr}
\toprule
  & \multicolumn{6}{c}{Predicted Pool} &  \multicolumn{6}{c}{Native Pool} & \\
\cmidrule{2-13}
Protein  & $E_{\mathit{None}}$  & $E_{\mathit{Acc}}$ & $E_{\mathit{Dist}}$ & $\sigma_{\mathit{Dist}}$ & $E_{\mathit{EPR}}$ & $\sigma_{\mathit{EPR}}$ & $E_{\mathit{None}}$ & $E_{\mathit{Acc}}$ & $E_{\mathit{Dist}}$ & $\sigma_{\mathit{Dist}}$ & $E_{\mathit{EPR}}$ & $\sigma_{\mathit{EPR}}$ \\
\midrule
1IWG   & \num{1.1}    & \num{1.5}  & \num{2.2}   & \num{0.2}   & \num{2.3}  & \num{0.2}  & \num{0.4}   & \num{0.9}  & \num{1.2}   & \num{0.4}   & \num{2.2}  & \num{0.3}  \\
1J4N   & \num{0.4}    & \num{0.3}  & \num{1.7}   & \num{0.2}   & \num{1.8}  & \num{0.2}  & \num{0.5}   & \num{0.6}  & \num{0.9}   & \num{0.2}   & \num{1.7}  & \num{0.3}  \\
1KPL   & \num{2.2}    & \num{2.4}  & \num{2.4}   & \num{0.1}   & \num{2.4}  & \num{0.3}  & \num{0.3}   & \num{0.3}  & \num{2.4}   & \num{0.4}   & \num{2.4}  & \num{0.2}  \\
1OCC   & \num{1.5}    & \num{1.5}  & \num{4.0}   & \num{1.0}   & \num{5.1}  & \num{1.0}  & \num{1.8}   & \num{1.8}  & \num{3.9}   & \num{1.0}   & \num{5.0}  & \num{1.0}  \\
1OKC   & \num{1.3}    & \num{1.4}  & \num{1.7}   & \num{0.2}   & \num{1.7}  & \num{0.2}  & \num{1.3}   & \num{1.4}  & \num{1.5}   & \num{0.1}   & \num{1.7}  & \num{0.2}  \\
1PV6   & \num{1.2}    & \num{1.3}  & \num{1.9}   & \num{0.2}   & \num{1.9}  & \num{0.3}  & \num{0.5}   & \num{1.0}  & \num{1.7}   & \num{0.4}   & \num{2.0}  & \num{0.3}  \\
1PY6   & \num{2.1}    & \num{2.0}  & \num{3.2}   & \num{0.3}   & \num{3.3}  & \num{0.3}  & \num{2.3}   & \num{2.3}  & \num{3.2}   & \num{0.3}   & \num{3.3}  & \num{0.3}  \\
1RHZ   & \num{1.1}    & \num{1.3}  & \num{1.8}   & \num{0.4}   & \num{2.1}  & \num{0.6}  & \num{1.3}   & \num{1.4}  & \num{1.6}   & \num{0.4}   & \num{1.4}  & \num{0.7}  \\
1U19   & \num{1.6}    & \num{2.2}  & \num{2.6}   & \num{0.2}   & \num{2.6}  & \num{0.2}  & \num{2.0}   & \num{1.7}  & \num{2.0}   & \num{0.3}   & \num{2.9}  & \num{0.7}  \\
1XME   & \num{1.4}    & \num{1.3}  & \num{1.7}   & \num{0.3}   & \num{1.7}  & \num{0.3}  & \num{1.1}   & \num{1.0}  & \num{1.7}   & \num{0.3}   & \num{3.6}  & \num{0.5}  \\
2BG9   & \num{0.9}    & \num{1.8}  & \num{2.5}   & \num{0.5}   & \num{2.5}  & \num{0.4}  & \num{1.6}   & \num{2.7}  & \num{1.6}   & \num{0.6}   & \num{1.9}  & \num{0.8}  \\
2BL2   & \num{0.5}    & \num{0.5}  & \num{1.4}   & \num{0.3}   & \num{1.4}  & \num{0.3}  & \num{1.1}   & \num{0.7}  & \num{3.1}   & \num{0.5}   & \num{4.9}  & \num{1.0}  \\
2BS2   & \num{2.0}    & \num{2.0}  & \num{2.6}   & \num{0.2}   & \num{2.6}  & \num{0.1}  & \num{1.9}   & \num{2.3}  & \num{2.4}   & \num{0.3}   & \num{2.8}  & \num{0.3}  \\
2IC8   & \num{1.0}    & \num{1.2}  & \num{1.6}   & \num{0.3}   & \num{1.7}  & \num{0.2}  & \num{1.2}   & \num{1.2}  & \num{1.7}   & \num{0.3}   & \num{2.8}  & \num{0.7}  \\
2K73   & \num{2.1}    & \num{1.5}  & \num{1.6}   & \num{0.6}   & \num{2.1}  & \num{0.2}  & \num{2.2}   & \num{2.4}  & \num{2.8}   & \num{0.6}   & \num{7.8}  & \num{0.7}  \\
2KSF   & \num{2.6}    & \num{2.5}  & \num{2.4}   & \num{0.8}   & \num{2.6}  & \num{0.7}  & \num{3.8}   & \num{3.2}  & \num{2.1}   & \num{0.7}   & \num{2.3}  & \num{0.7}  \\
2KSY   & \num{1.7}    & \num{2.2}  & \num{2.6}   & \num{0.7}   & \num{3.0}  & \num{0.6}  & \num{2.3}   & \num{2.0}  & \num{3.3}   & \num{0.6}   & \num{3.8}  & \num{1.1}  \\
2NR9   & \num{0.8}    & \num{0.9}  & \num{1.0}   & \num{0.2}   & \num{1.1}  & \num{0.2}  & \num{1.6}   & \num{1.2}  & \num{1.5}   & \num{0.3}   & \num{1.9}  & \num{0.2}  \\
2XUT   & \num{1.2}    & \num{1.3}  & \num{1.6}   & \num{0.2}   & \num{1.6}  & \num{0.2}  & \num{0.5}   & \num{0.5}  & \num{1.3}   & \num{0.2}   & \num{2.1}  & \num{0.3}  \\
3GIA   & \num{0.7}    & \num{0.7}  & \num{1.2}   & \num{0.2}   & \num{1.2}  & \num{0.2}  & \num{0.5}   & \num{0.6}  & \num{0.8}   & \num{0.2}   & \num{0.5}  & \num{0.1}  \\
3KCU   & \num{0.8}    & \num{1.5}  & \num{1.6}   & \num{0.2}   & \num{1.7}  & \num{0.2}  & \num{1.3}   & \num{1.1}  & \num{1.6}   & \num{0.2}   & \num{1.8}  & \num{0.5}  \\
3KJ6   & \num{1.6}    & \num{2.0}  & \num{2.3}   & \num{0.2}   & \num{2.2}  & \num{0.2}  & \num{1.4}   & \num{1.7}  & \num{2.0}   & \num{0.3}   & \num{4.2}  & \num{0.8}  \\
3P5N   & \num{0.8}    & \num{1.3}  & \num{1.5}   & \num{0.3}   & \num{1.6}  & \num{0.3}  & \num{1.3}   & \num{1.1}  & \num{1.7}   & \num{0.3}   & \num{2.3}  & \num{0.8}  \\
\midrule
2BHW   & \num{0.8}    & \num{0.9}  & \num{2.1}   & \num{0.6}   & \num{2.3}  & \num{0.5}  & \num{1.2}   & \num{1.7}  & \num{3.9}   & \num{1.2}   & \num{4.4}  & \num{1.7}  \\
2H8A   & \num{2.1}    & \num{2.4}  & \num{6.3}   & \num{0.4}   & \num{6.2}  & \num{0.7}  & \num{1.6}   & \num{2.1}  & \num{5.8}   & \num{0.9}   & \num{7.2}  & \num{0.9}  \\
2HAC   & \num{0.8}    & \num{2.8}  & \num{3.4}   & \num{0.6}   & \num{4.0}  & \num{0.9}  & \num{0.5}   & \num{2.4}  & \num{1.2}   & \num{0.2}   & \num{2.5}  & \num{0.7}  \\
2L35   & \num{0.8}    & \num{1.7}  & \num{1.9}   & \num{0.7}   & \num{1.9}  & \num{0.7}  & \num{0.8}   & \num{1.0}  & \num{2.3}   & \num{0.9}   & \num{2.6}  & \num{1.2}  \\
2ZY9   & \num{0.7}    & \num{2.3}  & \num{2.4}   & \num{0.2}   & \num{2.9}  & \num{0.3}  & \num{1.0}   & \num{1.4}  & \num{2.0}   & \num{0.4}   & \num{3.1}  & \num{0.4}  \\
3CAP   & \num{1.3}    & \num{1.8}  & \num{3.3}   & \num{0.4}   & \num{3.7}  & \num{0.4}  & \num{1.7}   & \num{2.2}  & \num{3.1}   & \num{0.8}   & \num{3.8}  & \num{0.2}  \\
\midrule
Ø      & \num{1.3}    & \num{1.6}  & \num{2.3}   & \num{0.3}   & \num{2.5}  & \num{0.4}  & \num{1.4}   & \num{1.5}  & \num{2.2}   & \num{0.5}   & \num{3.1}  & \num{0.6}  \\
\bottomrule
\end{tabular}

  \caption[Enrichments achieved for folding with and without EPR restraints]{\textbf{Enrichments
      achieved for folding with and without \gls{epr} restraints.} \Gls{epr} restraints
    significantly improve our ability to select the most accurate models among the sampled
    ones. When using \gls{epr} distance and accessibility restraints, the enrichment
    ($E_\mathit{EPR}$) could be improved in each case compared to structure prediction without
    \gls{epr} data ($E_\mathit{None}$). To be independent from specific spin labeling patterns ten
    different \gls{epr} distance restraint sets were used and the standard deviation regarding
    enrichment computed ($\sigma_\mathit{EPR}$). The experiment was also conducted using
    accessibility ($E_\mathit{Acc}$) and distance restraints ($E_\mathit{Dist}$ and
    $\sigma_\mathit{Dist}$) only. In addition to using predicted \glspl{sse} (Predicted pool), the
    experiment was repeated using \glspl{sse} obtained from the experimentally determined structure
    (Native pool). The proteins above the separating line are monomeric proteins; below the
    separating line are multimeric proteins.}
  \label{tab:mpepr_enrichment}
\end{sidewaystable}

\begin{sidewaystable}
  \small
  \centering
  \begin{tabular}{lrrrrrrrrrrrrrrrr}
\toprule
  &  \multicolumn{4}{c}{None} & \multicolumn{4}{c}{Accessibility} & \multicolumn{4}{c}{Distance} & \multicolumn{4}{c}{Accessibility \& Distance} \\
\cmidrule{2-17}
Protein  & $\mathit{best}$  & $\phi_{10}$  & $\gamma_{20}$  & $\gamma_{40}$  & $\mathit{best}$  & $\phi_{10}$  & $\gamma_{20}$  & $\gamma_{40}$  & $\mathit{best}$  & $\phi_{10}$  & $\gamma_{20}$ & $\gamma_{40}$  & $\mathit{best}$  & $\phi_{10}$  & $\gamma_{20}$  & $\gamma_{40}$  \\
\midrule
1IWG   & 34.5  & 30.5 & 3.3  & 0.0  & 44.8  & 42.2 & 19.2 & 0.3  & 43.5  & 37.0 & 7.4  & 0.0  & 48.9  & 44.5 & 22.0 & 1.0  \\
1J4N   & 38.8  & 35.3 & 7.8  & 0.0  & 51.0  & 42.9 & 26.5 & 0.2  & 47.8  & 42.8 & 17.5 & 0.2  & 51.6  & 46.0 & 29.0 & 0.5  \\
1KPL   & 18.4  & 15.1 & 0.0  & 0.0  & 21.7  & 17.8 & 0.0  & 0.0  & 23.0  & 18.9 & 0.0  & 0.0  & 20.7  & 17.7 & 0.0  & 0.0  \\
1OCC   & 44.0  & 38.5 & 4.7  & 0.1  & 65.7  & 61.9 & 32.2 & 2.9  & 65.5  & 59.6 & 19.9 & 2.4  & 74.3  & 68.5 & 32.5 & 7.5  \\
1OKC   & 15.1  & 13.2 & 0.0  & 0.0  & 18.9  & 15.8 & 0.0  & 0.0  & 20.9  & 17.4 & 0.0  & 0.0  & 20.2  & 17.9 & 0.0  & 0.0  \\
1PV6   & 18.7  & 17.5 & 0.0  & 0.0  & 27.1  & 23.8 & 0.3  & 0.0  & 28.2  & 22.6 & 0.2  & 0.0  & 28.6  & 23.1 & 0.2  & 0.0  \\
1PY6   & 23.2  & 22.0 & 0.2  & 0.0  & 49.1  & 36.1 & 4.0  & 0.0  & 36.0  & 30.4 & 2.8  & 0.0  & 38.2  & 33.6 & 4.2  & 0.0  \\
1RHZ   & 32.0  & 26.2 & 0.6  & 0.0  & 40.2  & 34.6 & 4.1  & 0.0  & 42.1  & 35.2 & 5.2  & 0.0  & 47.2  & 41.2 & 13.0 & 0.1  \\
1U19   & 25.0  & 20.5 & 0.1  & 0.0  & 29.9  & 27.7 & 2.3  & 0.0  & 30.1  & 25.6 & 0.9  & 0.0  & 40.2  & 34.0 & 3.9  & 0.0  \\
1XME   & 10.3  & 9.2  & 0.0  & 0.0  & 13.1  & 11.3 & 0.0  & 0.0  & 11.1  & 9.4  & 0.0  & 0.0  & 11.1  & 9.2  & 0.0  & 0.0  \\
2BG9   & 75.0  & 69.8 & 48.5 & 5.2  & 95.5  & 92.0 & 96.7 & 77.6 & 88.6  & 84.0 & 70.3 & 21.5 & 95.5  & 93.0 & 96.3 & 72.2 \\
2BL2   & 45.9  & 43.0 & 20.0 & 0.3  & 46.3  & 45.6 & 45.8 & 2.2  & 46.5  & 44.5 & 31.7 & 3.3  & 50.7  & 49.1 & 59.9 & 14.2 \\
2BS2   & 32.5  & 23.1 & 0.3  & 0.0  & 42.9  & 31.4 & 1.3  & 0.0  & 34.1  & 27.2 & 0.9  & 0.0  & 41.6  & 35.2 & 3.1  & 0.0  \\
2IC8   & 28.0  & 26.3 & 1.3  & 0.0  & 37.2  & 30.7 & 3.0  & 0.0  & 36.8  & 31.2 & 4.5  & 0.0  & 43.4  & 38.8 & 11.8 & 0.1  \\
2K73   & 53.5  & 45.6 & 7.7  & 0.2  & 69.8  & 65.5 & 45.6 & 5.8  & 63.0  & 55.7 & 25.3 & 2.4  & 74.2  & 70.2 & 54.8 & 13.3 \\
2KSF   & 45.3  & 40.8 & 6.4  & 0.1  & 84.9  & 76.6 & 43.5 & 7.5  & 67.0  & 60.5 & 29.7 & 2.8  & 85.7  & 81.3 & 61.8 & 16.1 \\
2KSY   & 29.2  & 24.5 & 0.5  & 0.0  & 38.8  & 35.3 & 6.6  & 0.0  & 37.8  & 33.2 & 3.6  & 0.0  & 49.2  & 43.7 & 10.5 & 0.3  \\
2NR9   & 62.3  & 55.2 & 10.7 & 0.9  & 77.9  & 73.6 & 56.0 & 15.9 & 61.7  & 56.8 & 43.1 & 4.3  & 68.6  & 65.4 & 90.1 & 33.3 \\
2XUT   & 75.0  & 69.8 & 48.5 & 5.2  & 21.7  & 19.8 & 0.1  & 0.0  & 21.9  & 20.0 & 0.1  & 0.0  & 24.4  & 21.6 & 0.3  & 0.0  \\
3GIA   & 14.4  & 10.5 & 0.0  & 0.0  & 12.1  & 10.9 & 0.0  & 0.0  & 12.0  & 10.3 & 0.0  & 0.0  & 11.2  & 10.2 & 0.0  & 0.0  \\
3KCU   & 8.9   & 7.6  & 0.0  & 0.0  & 10.0  & 8.5  & 0.0  & 0.0  & 9.2   & 7.6  & 0.0  & 0.0  & 8.8   & 7.3  & 0.0  & 0.0  \\
3KJ6   & 18.1  & 16.0 & 0.0  & 0.0  & 21.5  & 18.9 & 0.0  & 0.0  & 22.0  & 18.6 & 0.0  & 0.0  & 29.9  & 23.2 & 0.2  & 0.0  \\
3P5N   & 25.3  & 21.4 & 0.1  & 0.0  & 30.9  & 27.2 & 1.1  & 0.0  & 34.0  & 27.9 & 1.2  & 0.0  & 37.1  & 33.6 & 3.3  & 0.0  \\
\midrule
2BHW   & 25.7  & 23.0 & 0.4  & 0.0  & 33.5  & 27.5 & 1.5  & 0.0  & 30.7  & 27.1 & 1.6  & 0.0  & 36.0  & 31.9 & 3.7  & 0.0  \\
2H8A   & 100.0 & 99.7 & 85.2 & 77.1 & 100.0 & 99.8 & 89.0 & 81.5 & 100.0 & 99.6 & 86.1 & 79.0 & 100.0 & 99.8 & 91.7 & 83.5 \\
2HAC   & 49.2  & 45.6 & 8.7  & 0.3  & 56.5  & 45.9 & 7.5  & 0.3  & 51.6  & 45.1 & 7.2  & 0.2  & 66.3  & 57.3 & 8.1  & 1.4  \\
2L35   & 30.1  & 24.7 & 0.6  & 0.0  & 42.3  & 37.7 & 6.3  & 0.1  & 35.9  & 27.8 & 2.4  & 0.0  & 56.2  & 52.7 & 20.9 & 4.3  \\
2ZY9   & 19.8  & 17.1 & 0.0  & 0.0  & 27.7  & 23.1 & 0.3  & 0.0  & 23.5  & 19.0 & 0.0  & 0.0  & 29.6  & 25.8 & 0.8  & 0.0  \\
3CAP   & 20.9  & 17.7 & 0.0  & 0.0  & 26.7  & 21.7 & 0.2  & 0.0  & 22.2  & 18.8 & 0.0  & 0.0  & 26.9  & 22.0 & 0.1  & 0.0  \\
\midrule
Ø      & 35.1  & 31.3 & 8.8  & 3.1  & 42.7  & 38.1 & 17.0 & 6.7  & 39.5  & 35.0 & 12.5 & 4.0  & 45.4  & 41.3 & 21.5 & 8.5  \\
\bottomrule
\end{tabular}

  \caption[Prediction results when using secondary structure derived from the native
  structure]{\textbf{Prediction results when using secondary structure derived from the native
      structure.} Results for folding the proteins based on \glspl{sse} obtained from their native
    structure without restraints, with accessibility restraints, with distance restraints and with
    accessibility and distance restraints. Shown are the most accurate model sampled (best), the
    average of the ten most accurate models ($\mu_{10}$) as well as the percentage of the sampled
    models with \gls{rmsd100} values of less than \SI{4}{\angstrom} and \SI{8}{\angstrom} ($\tau_4$
    and $\tau_8$). The proteins above the separating line are monomeric proteins; below the
    separating line are multimeric proteins.}
  \label{tab:mpepr_contact_recovery_native}
\end{sidewaystable}

\begin{lstlisting}[breaklines=true, frame=single, caption=Selecting spin labeling sites for
\gls{sdsl}-\gls{epr} distance measurements. The first command optimizes the distributions of the
measurements using a \gls{mc} algorithm \autocite{Kazmier2011}. The second command simulates and
adds a \gls{cone} model-based uncertainty related to the translation from backbone distance into
spin-spin distance \autocite{Alexander2008}., label=lst:mpepr_dataset_generation]

bcl.exe OptimizeDataSetPairwise -fasta 1IWGA.fasta -pool_min_sse_lengths 0 0 -pool 1IWG.pool -distance_min_max 15 50 -nc_limit 10 -ensembles 1IWG_ensembles.ls -mc_number_iterations 100000 100000 -prefix 1IWG_ -nmodels 500 -read_scores_optimization opt_score_weights.wts -read_mutates_optimization mutate_weights.wts -data_set_size_fraction_of_sse_resis 0.2 -random_seed

bcl.exe SimulateDistanceRestraints -pdb 1IWGA.pdb -simulate_distance_restraints -output_file 1IWG.epr_cst_bcl -min_sse_size 0 0 0 -add_distance_uncertainty sl_cb.histograms -restraint_list 1IWG.epr 0 1 5 6 -random_seed
\end{lstlisting}

\begin{lstlisting}[breaklines=true, frame=single, caption=Creation of an \gls{sse} pool for the
protein 1IWG from \gls{sse} predictions from the method \gls{octopus} \autocite{Viklund2008}. The
``input'' folder must contain the fasta and \gls{octopus} prediction files for 1IWG.,
label=lst:mpepr_sse_generation]

bcl.exe CreateSSEPool -ssmethods OCTOPUS -pool_min_sse_lengths 5 3 -sse_threshold 0.5 0.5 0.5 -prefix 1IWG -join_separate -factory SSPredThreshold
\end{lstlisting}

\begin{lstlisting}[breaklines=true, frame=single, caption=Sample \num{40} models for the protein
1IWG from predicted \glspl{sse} using an \gls{mcm} algorithm. The \gls{sse} pool created in
\fref{lst:mpepr_sse_generation} is used as input data., label=lst:mpepr_structure_prediction]
bcl.exe protein:Fold -native 1IWGA.pdb -function_cache -pool_separate -min_sse_size 5 3 -quality RMSD GDT_TS -superimpose RMSD -sspred OCTOPUS JUFO9D -pool 1IWGA.SSPredHighest_JUFO9D_OCTOPUS.pool -stages_read stages.txt -pool_prefix 1IWGA -nmodels 40 -prefix 1IWG_dist_acc_pred_ -membrane -protein_storage pdbs/ -tm_helices 1IWGA.SSPredHighest_OCTOPUS.pool -sequence_data sspred/ 1IWG -opencl Disable -restraint_types DistanceEPR AccessibilityEPR -restraint_prefix restraints/1 -random_seed
\end{lstlisting}

\end{document}